\documentclass[10pt,journal,compsoc]{IEEEtran}
\usepackage[nocompress]{cite}
\usepackage{booktabs} 
\usepackage{amsmath,url,graphicx,amssymb}
\usepackage{amsfonts}
\usepackage{ctable}
\usepackage{multirow}
\usepackage{algorithm}
\usepackage{algpseudocode}
\usepackage{pifont}
\usepackage{color}
\usepackage{subfigure}
\usepackage{makecell}
\usepackage{bm}
\usepackage{enumitem}
\usepackage{caption}
\setlist{leftmargin=10pt}

\makeatletter
\def\blfootnote{\xdef\@thefnmark{}\@footnotetext}
\makeatother

\newtheorem{definition}{Definition}[section]

\newcommand{\mylistbegin}{
  \begin{list}{$\bullet$}
   {
     \setlength{\itemsep}{-2pt}
     \setlength{\leftmargin}{1em}
     \setlength{\labelwidth}{1em}
     \setlength{\labelsep}{0.5em} } }
\newcommand{\mylistend}{
   \end{list}  }

\newcommand{\eg}{\textit{e.g.}}

\newcommand{\aka}{\textit{a.k.a.}}
\newcommand{\ie}{\textit{i.e.}}
\newcommand{\etc}{\textit{etc}}

\newcommand{\header}[1]{{\vspace{+1mm}\flushleft \textbf{#1}}}

\newcommand{\bme}{{\bm e}}
\newcommand{\bmW}{{\bm W}}
\newcommand{\bmA}{{\bm A}}
\newcommand{\bmmu}{{\bm \mu}}
\newcommand{\bmg}{{\bm g}}
\newcommand{\bmx}{{\bm x}}
\newcommand{\bmy}{{\bm y}}
\newcommand{\bmw}{{\bm w}}

\newcommand{\bmb}{{\bm b}}
\newcommand{\bmh}{{\bm h}}

\begin{document}
\title{Heterogeneous Network Representation Learning: A Unified Framework with \\Survey and Benchmark}
\author{Carl Yang\IEEEauthorrefmark{1}\thanks{\IEEEauthorrefmark{1}Three authors contribute equally.}, Yuxin Xiao\IEEEauthorrefmark{1}, Yu Zhang\IEEEauthorrefmark{1},
        Yizhou Sun, \IEEEmembership{Member, IEEE,}
        and Jiawei Han, \IEEEmembership{Fellow, IEEE}
\IEEEcompsocitemizethanks{\IEEEcompsocthanksitem Carl Yang is with Emory University; Yuxin Xiao is with Carnegie Mellon University; Yu Zhang and Jiawei Han are with University of Illinois, Urbana Champaign; Yizhou Sun is with University of California, Los Angeles.\protect}
\thanks{Manuscript received Jun 2020; revised Oct 2020; accepted Dec 2020.}}

\markboth{IEEE Transactions on Knowledge and Data Engineering, ~Vol.~--, No.~--, Month~Year}
{Yang \MakeLowercase{\textit{et al.}}: Heterogeneous Network Representation Learning: A Unified Framework with Survey and Benchmark}

\IEEEtitleabstractindextext{
\begin{abstract}
Since real-world objects and their interactions are often multi-modal and multi-typed, heterogeneous networks have been widely used as a more powerful, realistic, and generic superclass of traditional
homogeneous networks (graphs). 
Meanwhile, representation learning (\aka~embedding) has recently been intensively studied and shown effective for various network mining and analytical tasks.
In this work, we aim to provide a unified framework to deeply summarize and evaluate existing research on heterogeneous network embedding (HNE), which includes but goes beyond a normal survey.
Since there has already been a broad body of HNE algorithms, as the first contribution of this work, we provide a generic paradigm for the systematic categorization and analysis over the merits of various existing HNE algorithms.
Moreover, existing HNE algorithms, though mostly claimed generic, are often evaluated on different datasets. Understandable due to the application favor of HNE, such indirect comparisons largely hinder the proper attribution of improved task performance towards effective data preprocessing and novel technical design, especially considering the various ways possible to construct a heterogeneous network from real-world application data. Therefore, as the second contribution, we create four benchmark datasets with various properties regarding scale, structure, attribute/label availability, and \etc.~from different sources, towards handy and fair evaluations of HNE algorithms. 
As the third contribution, we carefully refactor and amend the implementations and create friendly interfaces for 13 popular HNE algorithms, and provide all-around comparisons among them over multiple tasks and experimental settings. 

By putting all existing HNE algorithms under a unified framework, we aim to provide a universal reference and guideline for the understanding and development of HNE algorithms. Meanwhile, by open-sourcing all data and code, we envision to serve the community with an ready-to-use benchmark platform to test and compare the performance of existing and future HNE algorithms (https://github.com/yangji9181/HNE).
\end{abstract}

}

\maketitle
\IEEEdisplaynontitleabstractindextext
\IEEEpeerreviewmaketitle

\section{Introduction}
\label{sec:intro}
Networks and graphs constitute a canonical and ubiquitous paradigm for the modeling of interactive objects, which has drawn significant research attention from various scientific domains  \cite{seo2018structured, he2004manifold, gilmer2017neural, bordes2013translating, yang2017bridging, xiao2020discovering}. 
However, real-world objects and interactions are often multi-modal and multi-typed (\eg, authors, papers, venues and terms in a publication network \cite{sun2012mining, shi2016survey}; users, places, categories and GPS-coordinates in a location-based social network \cite{zhang2015geosoca, yang2019place, yang2018did}; and genes, proteins, diseases and species in a biomedical network \cite{li2010genome, davis2016comparative}). 
To capture and exploit such node and link heterogeneity, heterogeneous networks have been proposed and widely used in many real-world network mining scenarios, such as meta-path based similarity search \cite{sun2011pathsim, shi2014hetesim, yang2018similarity}, node classification and clustering \cite{dos2016multilabel, eswaran2017zoobp, chen2017task}, knowledge base completion \cite{socher2013reasoning, oh2018knowledge, zhang2019iteratively}, and recommendations \cite{geng2015learning, zhao2017meta, hou2017hindroid}.

In the meantime, current research on graphs has largely focused on representation learning (embedding), especially following the pioneer of neural network based algorithms that demonstrate revealing empirical evidence towards unprecedentedly effective yet efficient graph mining \cite{goyal2018graph, cai2018comprehensive, cui2018survey}. They aim to convert graph data (\eg, nodes \cite{perozzi2014deepwalk, tang2015line, grover2016node2vec, wang2018graphgan, kipf2016semi, hamilton2017inductive, chen2018fastgcn, velivckovic2018deep}, links \cite{zhao2016learning, abu2017learning, perozzi2017don, yang2020relation}, and subgraphs \cite{niepert2016learning, yang2018node, ying2018hierarchical, morris2019weisfeiler}) into low dimensional distributed vectors in the embedding space where the graph topological information (\eg, higher-order proximity \cite{cao2015grarep, wang2016structural, zhang2018arbitrary, huang2019hyper} and structure \cite{ribeiro2017struc2vec, zhang2017weisfeiler, lyu2017enhancing, donnat2018learning}) is preserved. Such \textit{embedding vectors} are then directly executable by various downstream machine learning algorithms \cite{scholkopf2001learning, liaw2002classification, chen2016xgboost}. 

Right on the intersection of heterogeneous networks and graph embedding, heterogeneous network embedding (HNE) recently has also received significant research attention \cite{chang2015heterogeneous, wu2015learning, zhao2015representation, dong2017metapath2vec, shi2018aspem, shi2018easing, gui2016large, fu2017hin2vec, yang2018meta, hussein2018meta, zhang2018deep, schlichtkrull2018modeling, qu2018curriculum, zhang2019shne, cen2019representation, zhang2019heterogeneous, hu2019adversarial, wang2019hyperbolic, yang2019neural, wang2019heterogeneous, lu2019relation}. Due to the application favor of HNE, many algorithms have been separately developed in different application domains such as search and recommendations \cite{geng2015learning, shi2018heterogeneous, cao2019unifying, yang2017bridging}. Moreover, as knowledge bases (KBs) also fall under the general umbrella of heterogeneous networks, many KB embedding algorithms can be compared with the HNE ones \cite{wang2017knowledge, bordes2013translating, lin2015learning, socher2013reasoning, yang2014embedding, dettmers2018convolutional, oh2018knowledge, wang2018knowledge, shang2019end}. 

Unfortunately, various HNE algorithms are developed in quite disparate communities across academia and industry.  They have never been systematically and comprehensively analyzed either in concepts or through experiments. In fact, due to the lack of benchmark platforms (with ready-to-use datasets and baselines), researchers often tend to construct their own datasets and re-implement a few most popular (sometimes outdated) algorithms for comparison, which renders fair performance evaluation and clear improvement attribution extremely hard, if not impossible. 

\begin{figure}[t]
\centering
\subfigure[Plain heterogeneous net.]{
\includegraphics[width=0.22\textwidth]{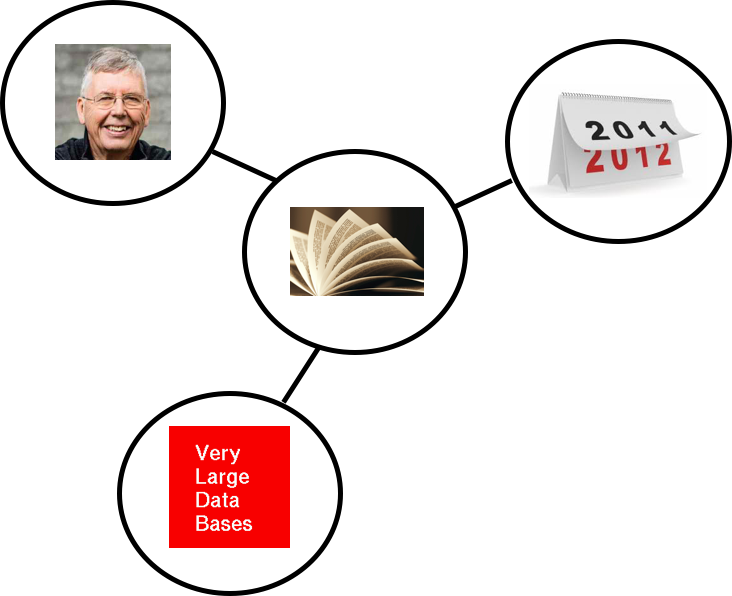}}
\hspace{5pt}
\subfigure[Labeled attributed net.]{
\includegraphics[width=0.22\textwidth]{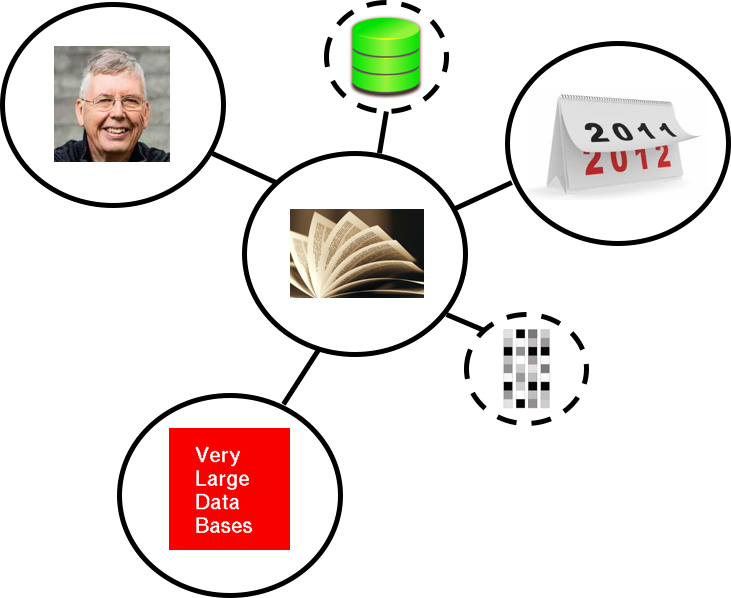}}
\caption{\textbf{Different heterogeneous networks constructed from the same real-world application data.}}
\label{fig:toy}
\end{figure}

Simply consider the toy examples of a publication dataset in Figure \ref{fig:toy}.\footnote{https://dblp.uni-trier.de/} Earlier HNE algorithms like metapath2vec \cite{dong2017metapath2vec} were developed on the heterogeneous network with node types of \textsf{authors, papers} and \textsf{venues} as in (a). However, one can enrich \textsf{papers} with a large number of \textsf{terms} and \textsf{topics} as additional nodes as in (b), which makes the random-walk based shallow embedding algorithms rather inefficient, but favors neighborhood aggregation based deep graph neural networks like R-GCN \cite{schlichtkrull2018modeling}. Moreover, one can further include node attributes like \textsf{term embedding} and labels like \textsf{research fields} and make them only available to the semi-supervised attributed embedding models, which may introduce even more bias \cite{zhang2018deep, wang2019heterogeneous, hu2020heterogeneous, ren2019heterogeneous}. Eventually, it is often hard to clearly attribute performance gains between technical novelty and data tweaking. 

In this work, we first formulate a unified yet flexible mathematical paradigm of HNE algorithms, easing the understanding of the critical merits of each model (Section \ref{sec:paradigm}). Particularly, based on a uniform taxonomy that clearly categorizes and summarizes the existing models (and likely future models), we propose a generic objective function of \textit{network smoothness}, and reformulate all existing models into this uniform paradigm while highlighting their individual novel contributions (Section \ref{sec:models}). We envision this paradigm to be helpful in guiding the development of future novel HNE algorithms, and in the meantime facilitate their conceptual contrast towards existing ones.

As the second contribution, we prepare four benchmark heterogeneous network datasets through exhaustive data collection, cleaning, analysis and curation (Section \ref{sec:data}).\footnote{https://github.com/yangji9181/HNE\label{repo}} The datasets we come up with cover a wide spectrum of application domains (\ie, publication, recommendation, knowledge base, and biomedicine), which have various properties regarding scale, structure, attribute/label availability, \etc.
This diverse set of data, together with a series of standard network mining tasks and evaluation metrics, constitute a handy and fair benchmark resource for future HNE algorithms.

As the third contribution, many existing HNE algorithms (including some very popular ones) either do not have a flexible implementation (\eg, hard-coded node and edge types, fixed set of meta-paths, \etc.), or do not scale to larger networks (\eg, high memory requirement during training), which adds much burden to novel research (\ie, requiring much engineering effort in correct re-implementation). To this end, we focus on 
13 popular HNE algorithms, where we carefully refactor and scale up the original implementations and apply additional interfaces for plug-and-run experiments on our prepared datasets (Section \ref{sec:exp}).\textsuperscript{\ref{repo}} Based on these ready-to-use and efficient implementations, we then conduct all-around empirical evaluations of the algorithms, and report their benchmark performances. The empirical results, while providing much insight into the merits of different models that are consistent with the conceptual analysis in Section \ref{sec:models}, also serve as the example utilization of our benchmark platform that can be followed by future studies on HNE.

Note that, although there have been several attempts to survey or benchmark heterogeneous network models \cite{sun2012mining, shi2016survey, wang2017knowledge, dong2020HeterNRL} and homogeneous graph embedding \cite{goyal2018graph, cai2018comprehensive, cui2018survey, wu2019comprehensive, dwivedi2020benchmarking, yue2020graph}, none of them has deeply looked into the intersection of the two. 
We advocate that our unified framework for the research and experiments on HNE is timely and necessary. Firstly, as we will cover in this work, there has been a significant amount of research on the particular problem of HNE especially in the very recent several years, but most of them scatter across different domains, lacking proper connections and comparisons. Secondly, none of the existing surveys has proposed a generic mathematically complete paradigm for conceptual analysis of all HNE models. Thirdly, existing surveys mostly do not provide systematic benchmark evaluation results, nor do they come with benchmark datasets and open-source baselines to facilitate future algorithm development.

The rest of this paper is organized as follows. Section \ref{sec:paradigm} first introduces our proposed generic HNE paradigm. Subsequently, representative models in our survey are conceptually categorized and analyzed in Section \ref{sec:models}. We then present in Section \ref{sec:data} our prepared benchmark datasets with detailed analysis. In Section \ref{sec:exp}, we provide a systematic empirical study over 13 popular HNE algorithms to benchmark the current state-of-the-art of HNE. Section \ref{sec:future} concludes the paper with visions towards future usage of our platform and research on HNE.

\section{Generic Paradigm}
\label{sec:paradigm}
\subsection{Problem Definitions}
\begin{definition}{Heterogeneous network.} 
A \textbf{heterogeneous network} $H=\{V, E, X, R, \phi, \psi\}$ is a network with multiple types of nodes and links. Particularly, within $H$, each node $v_i\in V$ is associated with a node type $\phi(v_i)$, and each link $e_{ij}\in E$ is associated with a link type $\psi(e_{ij})$. Each node $v_i$ of type $o=\phi(v_i)$ is also potentially associated with attribute $X^o_i$, while each link $e_{ij}$ of type $o=\psi(e_{ij})$ with attribute $R^o_{ij}$. It is worth noting that the type of a link $e_{ij}$ automatically defines the types of nodes $v_i$ and $v_j$ on its two ends.
\label{def:hin}
\end{definition}

\begin{figure}[t]
\centering
\includegraphics[width=0.45\textwidth]{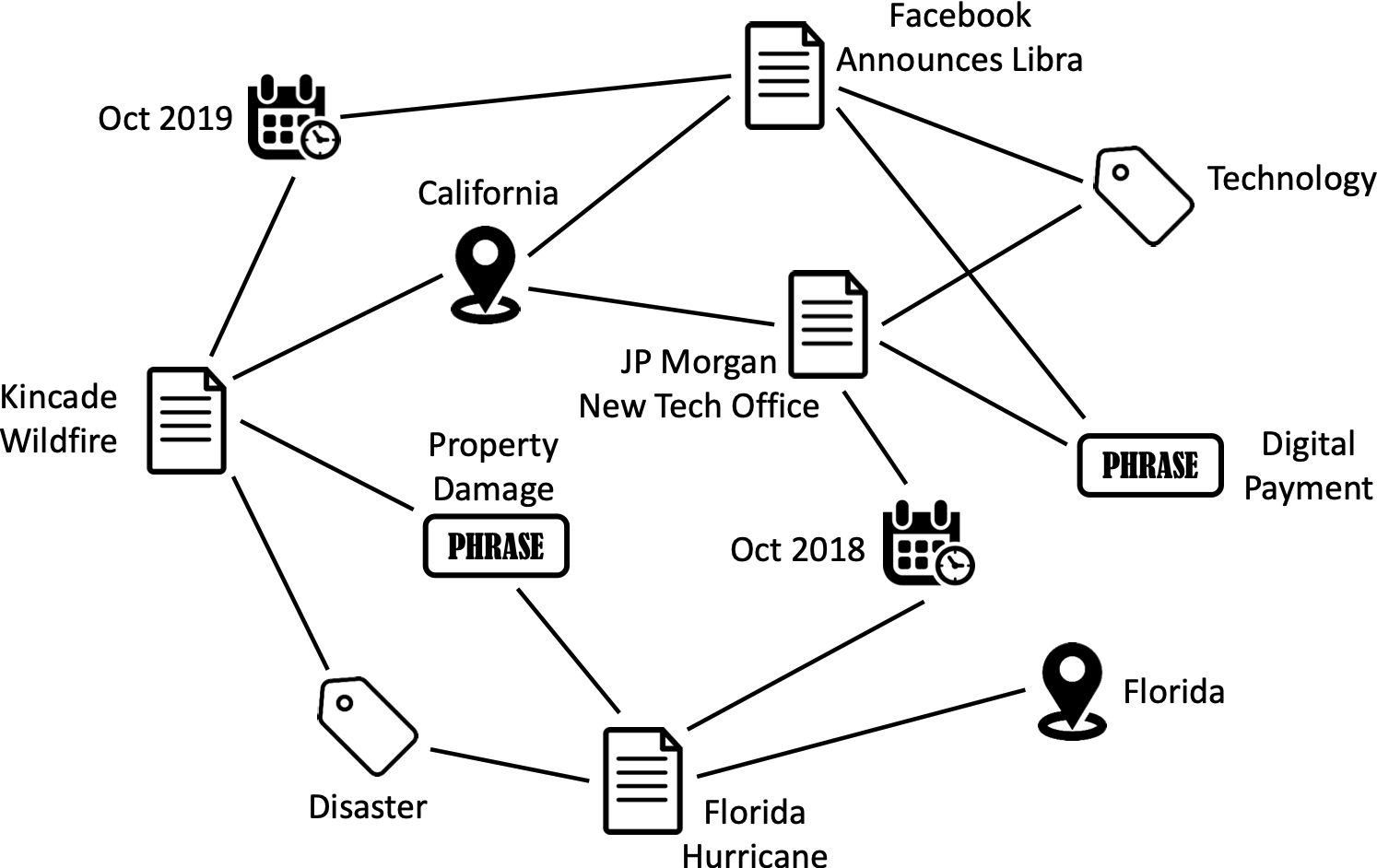}
\caption{\textbf{Toy example of a heterogeneous network constructed from the news data.}}
\label{fig:hin}
\end{figure}

Heterogeneous networks have been intensively studied due to its power of accommodating multi-modal multi-typed interconnected data. 
Besides the classic example of DBLP data used in most existing works as well as Figure \ref{fig:toy}, consider a different yet illustrative example from NYTimes in Figure \ref{fig:hin}.\footnote{https://www.nytimes.com/}  
Nodes in this heterogeneous network include \textsf{news articles}, \textsf{categories}, \textsf{phrases}, \textsf{locations}, and \textsf{datetimes}.
To illustrate the power of heterogeneous networks, we introduce the concept of meta-path, which has been leveraged by most existing works on heterogeneous network modeling \cite{sun2012mining, shi2016survey}.

\begin{definition}{Meta-path.}
A \textbf{meta-path} is a path defined on the network schema denoted in the form of $o_1\xrightarrow{l_1}o_2\xrightarrow{l_2}\cdots \xrightarrow{l_m}o_{m+1}$, where $o$ and $l$ are node types and link types, respectively.
\label{def:metapath}
\end{definition}

Each meta-path captures the proximity among the nodes on its two ends from a particular semantic perspective. Continue with our example of heterogeneous network from news data in Figure \ref{fig:hin}. The meta-path of \textsf{article} $\xrightarrow{\text{\sf belongs to}}$ \textsf{category} $\xrightarrow{\text{\sf includes}}$ \textsf{article} carries different semantics from \textsf{article} $\xrightarrow{\text{\sf mentions}}$ \textsf{location} $\xrightarrow{\text{\sf mentioned by}}$ \textsf{article}. Thus, the leverage of different meta-paths allows heterogeneous network models to compute the multi-modal multi-typed node proximity and relation, which has been shown beneficial to many real-world network mining applications \cite{shi2014hetesim, jiang2017semi, hou2017hindroid}.

Next, we introduce the problem of general network embedding (representation learning).
\begin{definition}{Network embedding.}
For a given network $G=\{V, E\}$, where $V$ is the set of nodes (vertices) and $E$ is the set of links (edges), a \textbf{network embedding} is a mapping function $\Phi\;:\;V\mapsto \mathbb{R}^{|V|\times d}$, where $d \ll |V|$. This mapping $\Phi$ defines the latent representation (\aka~embedding) of each node $v \in V$, which captures network topological information in $E$. 
\label{def:emb}
\end{definition}

In most cases, network proximity is the major topological information to be captured. For example, DeepWalk \cite{perozzi2014deepwalk} captures the random-walk based node proximity and illustrates the 2-dim node representations learned on the famous Zachary's Karate network of small groups, where a clear correspondence between the node position in the input graph and learned embedding space can be observed. Various follow-up works have improved or extended DeepWalk, while a complete coverage of them is beyond the scope of this work.
In this work, we focus on the embedding of heterogeneous networks.

Now we define the main problem of focus in this work, heterogeneous network embedding (HNE), which lies in the intersection between Def \ref{def:hin} and Def \ref{def:emb}.
\begin{definition}{Heterogeneous network embedding.}
For a given heterogeneous network $H$, a \textbf{heterogeneous network embedding} is a set of mapping functions $\{\Phi_k\;:\;V_k\mapsto \mathbb{R}^{|V_k|\times d}\}_{k=1}^K$, where $K$ is the number of node types, $\forall v_i\in V_k$, $\phi(v_i)=k,\; d \ll |V|$. Each mapping $\Phi_k$ defines the latent representation (\aka~embedding) of all nodes of type $k$, which captures the network topological information regarding the heterogeneous links in $E$. 
\label{def:hne}
\end{definition}

Compared with homogeneous networks, the definition of topological information in heterogeneous networks is even more diverse. As we will show in Section \ref{sec:models}, the major distinctions among different HNE algorithms mostly lie in their different ways of capturing such topological information. Particularly, the leverage of meta-paths as in Def \ref{def:metapath} often plays an essential role, since many popular HNE algorithms exactly aim to model the different proximity indicated by meta-paths \cite{dong2017metapath2vec, fu2017hin2vec, wang2019hyperbolic, cen2019representation, zhang2019shne, wang2019heterogeneous, zhang2018deep, shi2018heterogeneous}.

\subsection{Proposed Paradigm}
\label{sec:proposedpara}
In this work, we stress that one of the most important principles underlying HNE (as well as most other scenarios of network modeling and mining) is homophily \cite{mcpherson2001birds}. Particularly, in the network embedding setting, homophily can be translated as \textit{`nodes close on a network should have similar embeddings'}, which matches the requirement of Def \ref{def:emb}. In fact, we further find intrinsic connections between the well-perceived homophily principle and widely-used smoothness enforcement technique on networks \cite{belkin2002laplacian, zhou2004learning, zhu2003semi}, which leads to a generic mathematical paradigm covering most existing and likely many future HNE algorithms.

Based on earlier well-established concepts underlying network modeling and embedding learning \cite{tenenbaum2000global, roweis2000nonlinear, belkin2002laplacian, zhou2004learning, zhu2003semi}, we introduce the following key objective function of network smoothness enforcement as follows
\begin{align}
\mathcal{J} = \sum_{u, v \in V} w_{uv} d(\bme_u, \bme_v) + \mathcal{J}_R,
\label{eq:general}
\end{align}
where $\bme_u=\Phi(u)$ and $\bme_v=\Phi(v)$ are the node embedding vectors to be learned.  $w_{uv}$ is the proximity weight, $d(\cdot, \cdot)$ is the embedding distance function, and $\mathcal{J}_R$ denotes possible additional objectives such as regularizers, all three of which can be defined and implemented differently by the particular HNE algorithms.
\section{Algorithm Taxonomy}
\label{sec:models}
In this section, we find a universal taxonomy for existing HNE algorithms with three categories based on their common objectives, and elaborate in detail how they 
fit into our paradigm of Eq.~(\ref{eq:general}).
The main challenge of instantiating Eq.~(\ref{eq:general}) on heterogeneous networks is the consideration of complex interactions regarding multi-typed links and higher-order meta-paths. 
In fact, our Eq.~(\ref{eq:general}) also readily generalizes to homogeneous networks, though that is beyond the scope of this work.

\subsection{Proximity-Preserving Methods}

As mentioned above, one basic goal of network embedding is to capture network topological information. This can be achieved by preserving different types of proximity among nodes. There are two major categories of proximity-preserving methods in HNE: random walk approaches (inspired by DeepWalk \cite{perozzi2014deepwalk}) and first/second-order proximity based ones (inspired by LINE \cite{tang2015line}). Both types of proximity-preserving methods are considered as shallow network embedding, due to their essential single-layer decomposition of certain affinity matrices \cite{qiu2018network}.

\subsubsection{Random Walk Approaches}
\noindent \textbf{metapath2vec \cite{dong2017metapath2vec}.} Following homogeneous network embedding \cite{perozzi2014deepwalk,grover2016node2vec}, metapath2vec utilizes the node paths traversed by meta-path guided random walks to model the context of a node regarding heterogeneous semantics. Formally, given a meta-path $\mathcal{M} = o_1\xrightarrow{l_1}o_2\xrightarrow{l_2}\cdots \xrightarrow{l_{m-1}}o_{m}$, the transition probability at step $i$ is defined as

\begin{equation}
p(v_{i+1}|v_i, \mathcal{M}) =
\begin{cases}
\frac{1}{|\mathcal{N}_{l_i}(v_i)|} & \phi(v_{i+1}) = o_{i+1}, \psi(v_i, v_{i+1}) = l_i\\
0        & {\rm otherwise}
\end{cases}
\label{eqn:mp1}
\end{equation}
where $\mathcal{N}_l(v) = \{u|\psi(u,v) = l\}$ denotes the neighbors of $v$ associated with edge type $l$.
Assume $\mathcal{P} = \{\mathcal{P}_1,...,\mathcal{P}_M\}$ is the set of generated random walk sequences. The objective of metapath2vec is
\begin{equation}
\mathcal{J} = \sum_{v\in V}\sum_{u\in \mathcal{C}(v)} \log \frac{\exp(\bme_u^T\bme_v)}{\sum_{u'\in V}\exp(\bme_{u'}^T\bme_v)},
\label{eqn:mp2}
\end{equation}
where $\mathcal{C}(v)$ is the context (\ie, skip-grams) of $v$ in $\mathcal{P}$. For example, if $\mathcal{P}_1 = v_1v_2v_3v_4v_5...$ and the context window size is 2, then $\{v_1, v_2, v_4, v_5\} \subseteq \mathcal{C}(v_3)$. Let $w_{uv}$ be the number of times that $u \in \mathcal{C}(v)$, and we can rewrite Eq.~(\ref{eqn:mp2}) as
\begin{equation}
\mathcal{J} = \sum_{u, v\in V} w_{uv} \log \frac{\exp(\bme_u^T\bme_v)}{\sum_{u'\in V}\exp(\bme_{u'}^T\bme_v)}. \notag
\end{equation}
Calculating the denominator in this objective requires summing over all nodes, which is computationally expensive. In actual computation, it is approximated using negative sampling \cite{mikolov2013distributed,tang2015line}. 

\vspace{1mm}

\noindent \textbf{HIN2Vec \cite{fu2017hin2vec}.} HIN2Vec considers the probability that there is a meta-path $\mathcal{M}$ between nodes $u$ and $v$. Specifically,
\begin{equation}
p(\mathcal{M}|u,v) = \sigma\Big({\bf 1}^T\Big(\bmW_X^T{\bm u} \odot \bmW_Y^T{\bm v} \odot f_{01}\big(\bmW_R^T{\bm m}\big) \Big) \Big), \notag
\end{equation}
where $\bf 1$ is an all-ones vector; $\odot$ is the Hadamard product; $f_{01}$ is a normalization function. Here $\bme_u = \bmW_X^T{\bm u}$,  $\bme_v = \bmW_Y^T{\bm v}$ and $\bme_{\mathcal{M}} = f_{01}\big(\bmW_R^T{\bm m}\big)$ can be viewed as the embeddings of $u$, $v$ and $\mathcal{M}$, respectively. Let $\bmA_{\mathcal{M}} = diag(\bme_{\mathcal{M}1},...,$ $\bme_{\mathcal{M}d})$. We have
\begin{equation}
p(\mathcal{M}|u,v) = \sigma\Big({\bf 1}^T\big( \bme_u \odot \bme_v \odot \bme_\mathcal{M} \big) \Big) = \sigma(\bme_u^T\bmA_{\mathcal{M}}\bme_v). \notag
\end{equation}
$\sigma$ is the sigmoid function, so we have
\begin{equation}
1-p(\mathcal{M}|u,v) = 1- \sigma(\bme_u^T\bmA_{\mathcal{M}}\bme_v) = \sigma(-\bme_u^T\bmA_{\mathcal{M}}\bme_v). \notag
\end{equation}
HIN2Vec generates positive tuples $(u,v,\mathcal{M})$ (\ie, $u$ connects with $v$ via the meta-path $\mathcal{M}$) using homogeneous random walks \cite{perozzi2014deepwalk} regardless of node/link types. For each positive tuple $(u, v,\mathcal{M})$, it generates several negative tuples by replacing $u$ with a random node $u'$. Its objective is
\begin{equation}
\begin{split}
\mathcal{J}_0 &= \sum_{(u,v,\mathcal{M})} \log p(\mathcal{M}|u,v) + \sum_{(u',v,\mathcal{M})} \log (1-p(\mathcal{M}|u,v)) \\
&= \sum_{(u,v,\mathcal{M})} \Big(\log \sigma(\bme_u^T\bmA_{\mathcal{M}}\bme_v) + \sum_{u'} \log \sigma(-\bme_{u'}^T\bmA_{\mathcal{M}}\bme_v)\Big). \notag
\end{split}
\end{equation}
This is essentially the negative sampling approximation of the following objective
\begin{equation}
\begin{split}
\mathcal{J} &= \sum_{\mathcal{M}}\sum_{u, v\in V} w_{uv}^{(\mathcal{M})} \log \frac{\exp(\bme_u^T\bmA_{\mathcal{M}}\bme_v)}{\sum_{u'\in V}\exp(\bme_{u'}^T\bmA_{\mathcal{M}}\bme_v)}, \notag
\end{split}
\end{equation}
where $w_{uv}^{(\mathcal{M})}$ is the number of path instances between $u$ and $v$ following the meta-path $\mathcal{M}$. 



Other random walk approaches are summarized in Table \ref{tab:proximity}. To be specific, MRWNN \cite{wu2015learning} incorporates content priors into DeepWalk for image retrieval; SHNE \cite{zhang2019shne} incorporates additional node information like categorical attributes, images, \etc. by leveraging domain-specific deep encoders; HHNE \cite{wang2019hyperbolic} extends metapath2vec to the hyperbolic space; GHE \cite{chen2017task} proposes a semi-supervised meta-path weighting technique; MNE \cite{zhang2018scalable} conducts random walks separately for each view in a multi-view network; JUST \cite{hussein2018meta} proposes random walks with Jump and Stay strategies that do not rely on pre-defined meta-paths; HeteSpaceyWalk \cite{he2019hetespaceywalk} introduces a scalable embedding framework based on heterogeneous personalized spacey random walks; TapEm \cite{park2019task} proposes a task-guided node pair embedding approach for author identification.

\subsubsection{First/Second-Order Proximity Based Approaches}
\noindent \textbf{PTE \cite{tang2015pte}.} PTE proposes to decompose a heterogeneous network into multiple bipartite networks, each of which describes one edge type. Its objective is the sum of log-likelihoods over all bipartite networks: 
\begin{equation}
\begin{split}
\mathcal{J} &= \sum_{l \in \mathcal{T}_E} \sum_{u,v \in V} w_{uv}^{(l)} \log \frac{\exp(\bme_u^T\bme_v)}{\sum_{u'\in V_{\phi(u)}}\exp(\bme_{u'}^T\bme_v)} \\
&= \sum_{u,v \in V} w_{uv} \log \frac{\exp(\bme_u^T\bme_v)}{\sum_{u'\in V_{\phi(u)}}\exp(\bme_{u'}^T\bme_v)}. \notag
\end{split}
\end{equation}
Here $\mathcal{T}_E$ is the set of edge types; $w_{uv}^{(l)}$ is the type-$l$ edge weight of $(u,v)$ (if there is no edge between $u$ and $v$ with type $l$, then $w_{uv}^{(l)}=0$); $w_{uv}=\sum_{l}w_{uv}^{(l)}$ is the total edge weight between $u$ and $v$. 

\vspace{1mm}

\noindent \textbf{AspEm \cite{shi2018aspem}.} AspEm assumes that each heterogeneous network has multiple aspects, and each aspect is defined as a subgraph of the network schema \cite{sun2011pathsim}. An incompatibility measure is proposed to select appropriate aspects for embedding learning. Given an aspect $a$, its objective is
\begin{equation}
\mathcal{J} = \sum_{l \in \mathcal{T}_E^a} \frac{1}{Z_l}\sum_{u,v \in V} w_{uv}^{(l)} \log \frac{\exp(\bme_{u,a}^T \bme_{v,a})}{\sum_{u'\in V_{\phi(u)}}\exp(\bme_{u,a}^T \bme_{v,a})}, \notag
\end{equation}
where $\mathcal{T}_E^a$ is the set of edge types in aspect $a$; $Z_l = \sum_{u,v} w_{uv}^{(l)}$ is a normalization factor; $e_{u,a}$ is the aspect-specific embedding of $u$. 

\vspace{1mm}

\begin{table*}[h!]
	\centering
	\scalebox{0.97}{
		\begin{tabular}{c|c|c|c|c}
			\hline
			{\bf Algorithm} & $w_{uv}$ / $w_{uv}^{(l)}$ / $w_{uv}^{(\mathcal{M})}$ & $d(\bme_u,\bme_v)$ & $\mathcal{J}_{R0}$ &  {\bf Applications}\\
			\hline
			MRWNN \cite{wu2015learning} & \makecell{number of times that \\ $u \in \mathcal{C}(v)$ in homogeneous \\ random walks } & \multirow{2}{*}{$||\bme_u - \bme_v||^2$} & $-\sum_v ||\bme_v-f_{\rm ENC}(X_v)||^2$ & image retrieval\\
			\cline{1-2} \cline{4-5}
			metapath2vec \cite{dong2017metapath2vec} & \multirow{4}{*}{\begin{tabular}[c]{@{}c@{}} number of times that \\ $u \in \mathcal{C}(v)$ in heterogeneous \\ random walks following $\mathcal{M}$ \end{tabular}} &  & \multirow{16}{*}{N/A} & \multirow{4}{*}{\begin{tabular}[c]{@{}c@{}} node classification, \\ node clustering, \\ link prediction, \\ recommendation \end{tabular}} \\
			\cline{1-1} \cline{3-3} 
			SHNE \cite{zhang2019shne} & & \makecell{$||\bme_u - \bme_v||^2$, \\ $\bme_u = f_{\rm ENC}(\bmx_u)$} &  &  \\
			\cline{1-1} \cline{3-3} 
			HHNE \cite{wang2019hyperbolic} & & $d_{\mathbb{D}}(\bme_u, \bme_v)$ & & \\
			\cline{1-3} \cline{5-5}
			GHE \cite{chen2017task} &  \makecell{number of meta-path \\ instances between $u$ and $v$} & $||\bme_u - \bme_v||^2$ &   & author identification\\
			\cline{1-1} \cline{3-3} \cline{5-5}
			HIN2Vec \cite{fu2017hin2vec} &  following $\mathcal{M}$ & $||\sqrt{\bmA_\mathcal{M}}(\bme_u - \bme_v)||^2$ &  & \multirow{10}{*}{\begin{tabular}[c]{@{}c@{}} node classification, \\ node clustering, \\ link prediction, \\ recommendation \end{tabular}}\\
			\cline{1-3}
			MNE \cite{zhang2018scalable} & \makecell{number of times that \\ $u \in \mathcal{C}(v)$ in homogeneous \\ random walks in $(V, E_l)$} & $||\bme_u - \bme_{v,l}||^2$  & \\
			\cline{1-3}
			JUST \cite{hussein2018meta} & \makecell{number of times that \\ $u \in \mathcal{C}(v)$ in Jump/Stay \\ random walks \cite{hussein2018meta}} & \multirow{4}{*}{$||\bme_u-\bme_v||^2$}  &  & \\ 
			\cline{1-2}
			HeteSpaceyWalk \cite{he2019hetespaceywalk} & \makecell{number of times that \\ $u \in \mathcal{C}(v)$ in heterogeneous \\ spacey random walks \cite{he2019hetespaceywalk}} &  &  & \\ 
			\hline
			TapEm \cite{park2019task} & \makecell{number of times that \\ $u \in \mathcal{C}(v)$ in heterogeneous \\ random walks following $\mathcal{M}$} & \makecell{$||f_{\rm MAP}(\bme_u, \bme_v)-\bmh_{uv}^{\mathcal{M}}||^2$ \\ $\bmh_{uv}^{\mathcal{M}}$: representation of the \\ node sequence from $u$ to \\ $v$ following $\mathcal{M}$} & \makecell{supervised loss $\sum y\log\hat{y}$ \\ $+ (1-y)\log(1-\hat{y})$} & author identification\\ 
			\hline
			\hline
			HNE \cite{chang2015heterogeneous}  & ${\bf 1}_{(u,v) \in E}$ & \makecell{$||\bme_u-\bme_v||^2$, \\  $\bme_u = \bmA_{\phi(u)}\bmx_u$} & $-\sum_{o\in \mathcal{T}_V} ||\bmA_o||_F^2$
			&
			\makecell{text classification, \\image retrieval} \\
			\cline{1-5}
			PTE \cite{tang2015pte} & \multirow{4}{*}{edge weight of $(u,v)$} & \multirow{4}{*}{$||\bme_u-\bme_v||^2$} & \multirow{2}{*}{N/A} & text classification\\
			\cline{1-1} \cline{5-5}
			DRL \cite{qu2018curriculum} &  &  &  & node classification\\
			\cline{1-1} \cline{4-4} \cline{5-5}
			CMF \cite{zhao2015representation} & & & $-\sum_v ||\bme_v||^2$ & \makecell{relatedness measurement \\ of Wikipedia entities}  \\
			\hline
			HEBE \cite{gui2016large} & \makecell{edge weight of  \\ hyperedge $(u, C)$ \cite{gui2016large}} & \makecell{$||\bme_u-\bme_C||^2$, \\ $\bme_C = \sum_{v\in C}\bme_v/|C|$} & N/A & event embedding  \\
			\hline
			Phine \cite{yang2018did} & \makecell{number of times that $u$ and \\ $v$ co-occur in a meta-graph \\ instance } & $||\bme_u-\bme_v||^2$ & \makecell{supervised loss \\ $-\sum |\hat{y}-y|^2$} & user rating prediction \\
			\hline
			MVE \cite{qu2017attention} & \multirow{4}{*}{\begin{tabular}[c]{@{}c@{}} edge weight of $(u,v)$ \\ with type $l$ \end{tabular}} & $||\bme_{u,l}-\bme_v||^2$ & $-\sum_v\sum_l \lambda_{v,l}||\bme_{v,l} - \bme_v||^2 $ & \makecell{node classification, \\ link prediction} \\
			\cline{1-1} \cline{3-5}
			AspEm \cite{shi2018aspem} &  & $||\bme_{u,a}-\bme_{v,a}||^2$,\ \ $a$: aspect & \multirow{11}{*}{N/A} & aspect mining\\
			\cline{1-1} \cline{3-3} \cline{5-5}
			HEER \cite{shi2018easing} &  & $||\sqrt{\bmA_l}(\bme_u - \bme_v)||^2$ & & \makecell{relation prediction in \\ knowledge graphs} \\
			\cline{1-3} \cline{5-5}
			GERM \cite{jiang2020task} & \makecell{edge weight of $(u,v)$ with \\ type $l$ (if $l$ is activated by \\ the genetic algorithm \cite{jiang2020task})} & $||\bme_u - \bme_v||^2$ & & \makecell{citation recommendation \\ coauthor recommendation \\ feeds recommendation} \\
			\cline{1-3} \cline{5-5}
			mg2vec \cite{zhang2020mg2vec} & \makecell{1st: number of meta-graph \\ instances containing node $u$ \\ 2nd: number of meta-graph \\ instances containing both \\ nodes $u$ and $v$}  & \makecell{1st: $||\bme_v - \bme_m||^2$ \\ 2nd:\ \ \ \ \ \ \ \ \ \ \ \ \ \ \ \ \ \ \ \ \ \  \\ $||f_{\rm MAP}(\bme_u, \bme_v) - \bme_m||^2$ \\ $m$: meta-graph} & & \makecell{relationship prediction, \\ relationship search} \\
			\hline
		\end{tabular}
	}
	\caption{\label{tab:proximity} A summary of proximity-preserving based HNE algorithms. \textnormal{(Additional notations: $f_{\rm ENC}$: a function to encode text/image/attribute information; $d_{\mathbb{D}}$: distance between two points in the Poincar\'{e} ball; $f_{\rm MAP}$: a function to map two $d$-dimensional embeddings to one $d$-dimenional vector.)}}
	\vspace{-10pt}
\end{table*}

\noindent \textbf{HEER \cite{shi2018easing}.} HEER extends PTE by considering \textit{typed closeness}. Specifically, each edge type $l$ has an embedding $\bmmu_l$, and its objective is
\begin{equation}
\mathcal{J} = \sum_{l \in \mathcal{T}_E} \sum_{u,v \in V} w_{uv}^{(l)} \log \frac{\exp(\bmmu_l^T \bmg_{uv})}{\sum_{(u',v')\in P_l(u,v)}\exp(\bmmu_l^T \bmg_{u'v'})}, \notag
\end{equation}
where $\bmg_{uv}$ is the edge embedding of $(u,v)$; $P_l(u,v) = \{(u',v)|$ $\psi(u',v) = l\} \cup \{(u,v')|\psi(u,v') = l\}$. In \cite{shi2018easing}, $\bmg_{uv}$ has different definitions for directed and undirected edges based on the Hadamard product. To simplify our discussion, we assume $\bmg_{uv} = \bme_u \odot \bme_v$. Let $\bmA_l = diag(\bmmu_{l1},...,\bmmu_{ld})$. Then we have $\bmmu_l^T \bmg_{uv} = \bme_u^T \bmA_l \bme_v$ and
\begin{equation}
\mathcal{J} = \sum_{l \in \mathcal{T}_E} \sum_{u,v \in V} w_{uv}^{(l)} \log \frac{\exp(\bme_u^T \bmA_l \bme_v)}{\sum_{(u',v')\in P_l(u,v)}\exp(\bme_{u'}^T \bmA_l \bme_v)}. \notag
\end{equation}

Other first/second-order proximity based approaches are summarized in Table \ref{tab:proximity}. To be specific, 
Chang et al. \cite{chang2015heterogeneous} 
introduce a node type-aware content encoder; CMF \cite{zhao2015representation} performs joint matrix factorization over the decomposed bipartite networks; HEBE \cite{gui2016large} preserves proximity regarding each meta-graph; Phine \cite{yang2018did} combines additional regularization towards semi-supervised training; MVE \cite{qu2017attention} proposes an attenion based framework to consider multiple views of the same nodes; DRL \cite{qu2018curriculum} proposes to learn one type of edges at each step and uses deep reinforcement learning approaches to select the edge type for the next step; GERM \cite{jiang2020task} adopts a genetic evolutionary approach to preserve the critical edge-types while removing the noisy or incompatible ones given specific tasks; mg2vec \cite{zhang2020mg2vec} jointly embeds nodes and meta-graphs into the same space by exploiting both first-order and second-order proximity.

\subsubsection{Unified Objectives}
Based on the discussions above, the objective of most 
proximity-preserving methods can be unified as
\begin{equation}
\max \mathcal{J} = \sum_{u,v} w_{uv} \log \frac{\exp(s(u,v))}{\sum_{u'} \exp(s(u',v))} + \mathcal{J}_{R0}.
\label{eqn:obj}
\end{equation}
Here, $w_{uv}$ is the weight of node pair $(u,v)$ in the objective\footnote{$w_{uv}$ can be specific to a meta-path $\mathcal{M}$ or an edge type $l$, in which cases we can denote it as $w_{uv}^{(\mathcal{M})}$ or $w_{uv}^{(l)}$ accordingly. In Eq.~(\ref{eqn:obj}) and following derivations, for ease of notation, we omit the superscript.}; $\mathcal{J}_{R0}$ is an algorithm-specific regularization term (summarized in Table \ref{tab:proximity}); $s(u,v)$ is a proximity function between $u$ and $v$.

\vspace{1mm}

\noindent \textbf{Negative Sampling.} Directly optimizing the objective in Eq.~(\ref{eqn:obj}) is computationally expensive because it needs to traverse all nodes when computing
the softmax function. Therefore, in actual computation, most studies adopt the negative sampling strategy \cite{mikolov2013distributed,tang2015line}, which modifies the objective as follows:
\begin{equation}
    \sum_{u,v}w_{uv}\Big(\log\sigma(s(u, v)) + b\mathbb{E}_{u'\sim P_N}[\log\sigma(s(u', v))]\Big) + \mathcal{J}_{R0}. \notag
\end{equation}
Here, $b$ is the number of negative samples; $P_N$ is known as the noise distribution that generates negative samples.

Negative sampling serves as a generic paradigm to unify network embedding approaches. For example, starting from the negative sampling objective, Qiu et al. \cite{qiu2018network} unify DeepWalk \cite{perozzi2014deepwalk}, LINE \cite{tang2015line}, PTE \cite{tang2015pte} and node2vec \cite{grover2016node2vec} into a matrix factorization framework. In this paper, as mentioned in Section \ref{sec:proposedpara}, we introduce another objective (\ie, Eq.~(\ref{eq:general})) that is equivalent to Eq.~(\ref{eqn:obj}) and consider network embedding from the perspective of network smoothness enforcement.

\vspace{1mm}

\noindent \textbf{Network Smoothness Enforcement.} Note that in most cases, we can write $s(u,v)$ in Eq.~(\ref{eqn:obj}) as $f(\bme_u)^T f(\bme_v)$. For example, $f(\bme_u) = \bme_u$ in metapath2vec, PTE, \etc.; $f(\bme_u) = \sqrt{\bmA_\mathcal{M}}\bme_u$ in HIN2Vec;  $f(\bme_u) = \sqrt{\bmA_l}\bme_u$ in HEER. In these cases,
\begin{equation}
\begin{split}
\mathcal{J} &= \sum_{u,v} w_{uv}s(u,v) -\sum_{u,v} w_{uv}\log\sum_{u'} \exp(s(u',v)) + \mathcal{J}_{R0} \\
&= \sum_{u,v} w_{uv} f(\bme_u)^T f(\bme_v) \\
& \ \ \ \ - \sum_{u,v} w_{uv}\log\sum_{u'} \exp(f(\bme_{u'})^T f(\bme_v)) + \mathcal{J}_{R0} \\
&= \sum_{u,v} \frac{w_{uv}}{2}\Big(||f(\bme_u)||^2 + ||f(\bme_v)||^2 -||f(\bme_u) - f(\bme_v)||^2 \Big) \\
& \ \ \ \ - \sum_{u,v} w_{uv}\log\sum_{u'} \exp(f(\bme_{u'})^T f(\bme_v)) + \mathcal{J}_{R0}. \notag
\end{split}
\end{equation}
The last step holds because $||\bmx-\bmy||^2 = (\bmx-\bmy)^T(\bmx-\bmy) = ||\bmx||^2+||\bmy||^2-2\bmx^T\bmy$. Therefore, our goal is equivalent to
\begin{equation}
\begin{split}
\min -\mathcal{J} &= \sum_{u,v} \frac{w_{uv}}{2}\underbrace{||f(\bme_u) - f(\bme_v)||^2}_{d(\bme_u,\bme_v)} - \mathcal{J}_{R0} \\
& - \underbrace{\sum_{u,v} \frac{w_{uv}}{2} \Big(||f(\bme_u)||^2 + ||f(\bme_v)||^2\Big)}_{\mathcal{J}_{R1}} \\
& + \underbrace{\sum_{u,v} w_{uv} \log \sum_{u'} \exp(f(\bme_{u'})^T f(\bme_v))}_{\mathcal{J}_{R2}}. 
\label{eq:ppj}
\end{split}
\end{equation}
Here $\mathcal{J}_{R1}$ and $\mathcal{J}_{R2}$ are two regularization terms. Without $\mathcal{J}_{R1}$, $d(\bme_u,\bme_v)$ can be minimized by letting $||f(\bme_u)||\rightarrow 0 \ (\forall u \in V)$; without $\mathcal{J}_{R2}$, $d(\bme_u,\bme_v)$ can be minimized by letting $ \bme_u \equiv {\bm c}\ (\forall u \in V)$. $\mathcal{J}_R$ in Eq.~(\ref{eq:general}) is then the sum of $-\mathcal{J}_{R0}$, $-\mathcal{J}_{R1}$ and $\mathcal{J}_{R2}$. 
Among them, $-\mathcal{J}_{R0}$ is algorithm-specific, while $-\mathcal{J}_{R1}$ and $\mathcal{J}_{R2}$ are commonly shared across most HNE algorithms in the proximity-preserving group. We summarize the choices of $w_{uv}$, $d(e_u, e_v)$ and $\mathcal{J}_{R0}$ in Table \ref{tab:proximity}. 

Although Eq.~(\ref{eqn:obj}) can cover most existing and likely many future approaches, we would like to remark that there are also studies adopting other forms of proximity-preserving objectives. For example, SHINE \cite{wang2018shine} uses reconstruction loss of autoencoders; HINSE \cite{yang2018meta} adopts spectral embedding based on adjacency matrices with different meta-graphs; MetaDynaMix \cite{fard2019relationship} introduces a low-rank matrix factorization framework for dynamic HIN embedding; HeGAN \cite{hu2019adversarial} proposes an adversarial learning approach with a relation type-aware discriminator.

\subsection{Message-Passing Methods}

Each node in a network can have attribute information represented as a feature vector $\bmx_u$. Message-passing methods aim to learn node embeddings $\bme_u$ based on $\bmx_u$ by aggregating the information from $u$'s neighbors. In recent studies, Graph Neural Networks (GNNs) \cite{kipf2016semi} are widely adopted to facilitate this aggregation/message-passing process. Compared to the proximity based HNE methods, message-passing methods, especially the GNN based ones are often considered as deep network embedding, due to their multiple layers of learnable projection functions.

\vspace{1mm}
 
\noindent \textbf{R-GCN \cite{schlichtkrull2018modeling}.} R-GCN has $K$ convolutional layers. The initial node representation $\bmh_u^{(0)}$ is just the node feature $\bmx_u$. In the $k$-th convolutional layer, each representation vector is updated by accumulating the vectors of neighboring nodes through a normalized sum.
\begin{equation}
\bmh_u^{(k+1)} = \sigma\Big(\sum_{l\in \mathcal{T}_E} \sum_{v\in \mathcal{N}_l(u)} \frac{1}{|\mathcal{N}_l(u)|} \bmW_l^{(k)}\bmh_v^{(k)}+\bmW_0^{(k)}\bmh_u^{(k)}\Big). \notag
\end{equation}
Different from the regular GCN model \cite{kipf2016semi}, R-GCN considers edge heterogeneity by learning multiple convolution matrices $\bmW$'s, each of which corresponding to one edge type. During message passing, neighbors under the same edge type will be aggregated and normalized first. The node embedding is the output of the $K$-th layer (\ie, $\bme_v = \bmh_v^{(K)}$).

In unsupervised settings, message-passing approaches use link prediction as their downstream task to train GNNs. To be specific, the likelihood of observing edges in the heterogeneous network is maximized. R-GCN optimizes a cross-entropy loss through negative sampling. Essentially, it is the approximation of the following objective:
\begin{equation}
\mathcal{J} = \sum_{l \in \mathcal{T}_E} \sum_{u,v \in V} w_{uv}^{(l)} \log \frac{\exp(\bme_u^T\bmA_l\bme_v)}{\sum_{u'\in V}\exp(\bme_u^T\bmA_l\bme_v)}, \notag
\end{equation}
where $w_{uv}^{(l)} = \textbf{1}_{(u,v)\in E_l}$.

Recently, CompGCN \cite{vashishth2019composition} extends R-GCN by leveraging a variety of entity-relation composition operations so as to jointly embed nodes and relations.

\begin{table*}[h!]
\centering
\scalebox{0.93}{
\begin{tabular}{c|c|c|c|c}
	\hline
	{\bf Algorithm} & $w_{uv}$ / $w_{uv}^{(l)}$ / $w_{uv}^\mathcal{M}$ & $d(\bme_u, \bme_v)$ & Aggregation Function & {\bf Applications}\\
	\hline
	R-GCN \cite{schlichtkrull2018modeling}  &
	$\textbf{1}_{(u,v)\in E_l}$ &
	$||\sqrt{\bmA_l}(\bme_u - \bme_v)||^2$ &
	\makecell{$\bmh_u^{(k+1)} = \sigma\Big(\sum_{l\in \mathcal{T}_E} \sum_{v \in \mathcal{N}_l(u)} \frac{1}{|\mathcal{N}_l(u)|} \bmW_r^{(k)}\bmh_v^{(k)}+\bmW_0^{(k)}\bmh_u^{(k)}\Big)$ \\ $\bmh_u^{(0)} = \bmx_u$, \ \ $\bme_u = \bmh_u^{(K)}$} &
	\makecell{entity classification, \\KB completion} \\
	\hline
	HEP \cite{zheng2018heterogeneous} &
	$\textbf{1}_{(u,v)\in E}$ &
	$||\sqrt{\bmA}(\bme_u - \bme_v)||^2$ & $\tilde{\bme}_u = \sigma\Big(\bmW_{\phi(u)}\Big({\big|\big|}_{o\in \mathcal{T}_V} \sum_{v\in \mathcal{N}_o(u)} \alpha_{uv}\bme_v\Big)+\bmb_{\phi(u)}\Big)$ & user alignment  \\
	\hline
	HAN \cite{wang2019heterogeneous} & \makecell{edge weight \\ of $(u,v)$} & \multirow{5}{*}{$||\bme_u-\bme_v||^2$} & $\bme_u = \sum_{\mathcal{M}}\beta_{\mathcal{M}} \sigma\Big(\sum_{v\in \mathcal{N}_{\mathcal{M}}(u)} \alpha_{uv}^{\mathcal{M}}{\bm M}_{\phi(u)}\bmx_u\Big)$ & \multirow{14}{*}{\begin{tabular}[c]{@{}c@{}} node classification, \\ node clustering, \\ link prediction, \\ recommendation \end{tabular}}  \\
	\cline{1-2} \cline{4-4}
	HetGNN \cite{zhang2019heterogeneous} & \makecell{number of times \\ that $u \in \mathcal{C}(v)$ in \\ homogeneous \\ random walks} &  & \makecell{$\bmh_u = f_{\rm ENC}(\bmx_u), \ \ \bmh_{u,o} = f_{\rm AGG}\Big(\{\bmh_v| v\in \mathcal{N}_{\rm RWR}(u), \phi(v) = o\}\Big)$ \\ $\bme_u = \alpha_u \bmh_u + \sum_{o \in \mathcal{T}_V} \alpha_{u,o} \bmh_{u,o}$} \\
	\cline{1-4}
	GATNE \cite{cen2019representation} & \multirow{4}{*}{\begin{tabular}[c]{@{}c@{}} number of times \\ that $u \in \mathcal{C}(v)$ in \\ random walks \\ following $\mathcal{M}$ \end{tabular}} & \multirow{4}{*}{$||\bme_{u,l} - \bme_{v}||^2$} & \makecell{$\bmh_{u,l}^{(k+1)} = f_{\rm AGG}\Big(\{\bmh_{v,l}^{(k)}|v\in \mathcal{N}_l(u)\}\Big)$, \ \ $\bmh_{u,l}^{(0)} = \bmx_u$ \\ $\bme_{u,l} = \alpha_l\bmW_l\Big({\big|\big|}_{l\in \mathcal{T}_E}\bmh_{u,l}^{(K)}\Big){\bm a}_{u,l} + \bmb_u$} &  \\
	\cline{1-1} \cline{4-4}
	MV-ACM \cite{zhao2020deep} &  &  & \makecell{$\bmh_{u,l}^{(k+1)} = \sigma\Big(\bmW_{l}^{(k)}\frac{1}{|\mathcal{N}_l(u)|}\sum_{v\in \mathcal{N}_l(u)}\bmh_{v,l}^{(k)}\Big)$, \ \ $\bmh_{u,l}^{(0)} = \bmx_u$ \\ $\bme_{u,l} = {\bm M_l}\Big( \bmh_{u,l}^{(K)} +\sum_{l'\in \mathcal{T}_E} \alpha_{u,l,l'} \bmh_{u,l'}^{(K)} \Big) + \bmb_u$} & \\ 
	\cline{1-4}
	MAGNN \cite{fu2020magnn} & \makecell{edge weight \\ of $(u,v)$} & $||\bme_u - \bme_{v}||^2$ & \makecell{$\bmh_u^{\mathcal{M}} = \sigma\Big(\sum_{v\in \mathcal{N}_{\mathcal{M}}(u)} \alpha_{uv}^{\mathcal{M}} f_{\rm ENC}\big(\{{\bm M}_{\phi(t)}\bmx_t|t\in \mathcal{P}_{u\rightarrow v}^{\mathcal{M}}\}\big)\Big)$ \\ $\bme_u = \sigma\Big(\bmW \big(\sum_{\mathcal{M}}\beta_{\mathcal{M}} \bmh_u^{\mathcal{M}}\big)\Big)$} &  \\
	\hline
	HGT \cite{hu2020heterogeneous} & $\textbf{1}_{(u,v)\in E}$ & \makecell{refer to NTN \cite{socher2013reasoning} \\ in Table \ref{tab:translation}} & \makecell{$\hat{\bmh}^{(k+1)}_v = \sum_{u\in N(v)} {\bf Attention}(u,v)\odot {\bf Message}(u,v)$ \\ $\bmh^{(k+1)}_v = A_{\phi(v)}(\sigma(\hat{\bmh}^{(k+1)}_v))+ \bmh^{(k)}_v$,\ \  $\bmh_u^{(0)} = \bmx_u$, \ \ $\bme_u = \bmh_u^{(K)}$} & \makecell{node classification, \\ author identification} \\ 
	\hline
\end{tabular}
}
\caption{\label{tab:message} A summary of message-passing based HNE algorithms. \textnormal{(Additional notations: $\mathcal{N}_{\rm RWR}(v)$: neighbors of $v$ defined through random walk with restart \cite{tong2006fast}; $\mathcal{N}_l(u)$: neighbors of $u$ with edge type $l$; $\mathcal{N}_o(u)$: neighbors of $u$ with node type $o$; $\mathcal{N}_{\mathcal{M}}(u)$: nodes connects with $u$ via meta-path $\mathcal{M}$; $\mathcal{P}_{u\rightarrow v}^{\mathcal{M}}$: a meta-path instance of $\mathcal{M}$ connecting $u$ and $v$; $f_{\rm AGG}$: a function to aggregate information from neighbors. We show the transductive version of GATNE.)}} 
\end{table*}

\vspace{1mm}

\noindent \textbf{HAN \cite{wang2019heterogeneous}.} Instead of considering one-hop neighbors, HAN utilizes meta-paths to model higher-order proximity. Given a meta-path $\mathcal{M}$, the representation of node $u$ is aggregated from its meta-path based neighbors $\mathcal{N}_{\mathcal{M}}(u) = \{u\}\cup\{v|v$ connects with $u$ via the meta-path $\mathcal{M}\}$. HAN proposes an attention mechanism to learn the weights of different neighbors:
\begin{gather}
\alpha_{uv}^{\mathcal{M}} = \frac{\exp\big(\sigma({\bm a}_{\mathcal{M}}^T[\bmx'_u || \bmx'_v])\big)}{\sum_{v'\in \mathcal{N}_{\mathcal{M}}(u)}\exp\big(\sigma({\bm a}_{\mathcal{M}}^T[\bmx'_u || \bmx'_{v'}])\big)}, \notag \\
\bmh_u^{\mathcal{M}} = \sigma\Big(\sum_{v\in \mathcal{N}_{\mathcal{M}}(u)} \alpha_{uv}^{\mathcal{M}} \bmx'_v \Big), \notag
\end{gather}
where ${\bm a}_{\mathcal{M}}$ is the node-level attention vector of $\mathcal{M}$; $\bmx'_u = {\bm M}_{\phi(u)}\bmx_u$ is the projected feature vector of node $u$; $||$ is the concatenation operator. Given the meta-path specific embedding $\bmh_u^{\mathcal{M}}$, HAN uses a semantic-level attention to weigh different meta-paths:
\begin{gather}
\beta_{\mathcal{M}} = \frac{\exp\big(\frac{1}{|V|}\sum_{v\in V}{\bm q}^T \text{tanh}(\bmW \bmh_v^{\mathcal{M}}+\bmb)\big)}{\sum_{\mathcal{M}'} \exp\big(\frac{1}{|V|}\sum_{v\in V}{\bm q}^T \text{tanh}(\bmW \bmh_v^{\mathcal{M'}}+\bmb)\big)}, \notag \\
\bme_u = \sum_{\mathcal{M}}\beta_{\mathcal{M}} \bmh_u^{\mathcal{M}}, \notag 
\end{gather}
where ${\bm q}$ is the semantic-level attention vector.

In the original HAN paper, the authors mainly consider the task of semi-supervised node classification. For unsupervised learning (\ie, without any node labels), according to \cite{fu2020magnn}, HAN can use the link prediction loss introduced in GraphSAGE \cite{hamilton2017inductive}, which is the negative sampling approximation of the following objective:
\begin{equation}
\mathcal{J} = \sum_{u,v \in V} w_{uv} \log \frac{\exp(\bme_u^T\bme_v)}{\sum_{u'\in V} \exp(\bme_{u'}^T\bme_v)}.
\label{eqn:graphsage}
\end{equation}
Here, $w_{uv}$ is the edge weight of $(u, v)$.



\vspace{1mm}

\noindent \textbf{MAGNN \cite{fu2020magnn}.} MAGNN extends HAN by considering both the meta-path based neighborhood $\mathcal{N}_{\mathcal{M}}(u) = \{v|v$ connects with $u$ via meta-path $\mathcal{M}\}$ and the nodes along the meta-path instances. Given a meta-path $\mathcal{M}$, MAGNN first employs an encoder to transform all the node features along an instance of $\mathcal{M}$ into a single vector.
\begin{equation}
    \bmh_{uv}^{\mathcal{M}} = f_{\rm ENC}\big(\{\bmx'_t|t\in \mathcal{P}_{u\rightarrow v}^{\mathcal{M}}\}\big), \notag
\end{equation}
where $\bmx'_u = {\bm M}_{\phi(u)}\bmx_u$ is the projected feature vector of node $u$; $\mathcal{P}_{u\rightarrow v}^{\mathcal{M}}$ denotes a meta-path instance of $\mathcal{M}$ connecting $u$ and $v$; $f_{\rm ENC}$ is a relational rotation encoder inspired by \cite{sun2019rotate}. After encoding each meta-path instance, MAGNN proposes an intra-meta-path aggregation to learn the weight of different neighbors:
\begin{gather}
\alpha_{uv}^{\mathcal{M}} = \frac{\exp(\text{LeakyReLU}({\bm a}_{\mathcal{M}}^T[\bmx'_v||\bmh_{uv}^{\mathcal{M}}]))}{\sum_{v'\in \mathcal{N}_{\mathcal{M}}(u)} \exp(\text{LeakyReLU}({\bm a}_{\mathcal{M}}^T[\bmx'_{v'}||\bmh_{uv'}^{\mathcal{M}}]))}, \notag \\
\bmh_u^{\mathcal{M}} = \sigma\Big(\sum_{v\in \mathcal{N}_{\mathcal{M}}(u)} \alpha_{uv}^{\mathcal{M}} \bmh_{uv}^{\mathcal{M}} \Big), \notag
\end{gather}
where ${\bm a}_{\mathcal{M}}$ is the node-level attention vector of $\mathcal{M}$. This attention mechanism can also be extended to multiple heads. After aggregating the information within each meta-path, MAGNN further combines the semantics revealed by all meta-paths using an inter-meta-path aggregation:
\begin{equation}
\footnotesize
\begin{split}
\beta_{\mathcal{M}} = \frac{\exp\big(\frac{1}{|V_{\phi(u)}|}\sum_{v \in V_{\phi(u)}}{\bm q}_{\phi(u)}^T{\rm tanh}(\bmW_{\phi(u)} \bmh_v^{\mathcal{M}}+\bmb_{\phi(u)})\big)}{\sum_{\mathcal{M}'} \exp\big(\frac{1}{|V_{\phi(u)}|} \sum_{v \in V_{\phi(u)}}{\bm q}_{\phi(u)}^T{\rm tanh}(\bmW_{\phi(u)} \bmh_v^{\mathcal{M'}}+\bmb_{\phi(u)})\big)}, \notag
\end{split}
\end{equation}
\begin{equation}
\bme_u = \sigma\Big(\bmW \big(\sum_{\mathcal{M}}\beta_{\mathcal{M}} \bmh_u^{\mathcal{M}}\big)\Big), \notag 
\end{equation}
where ${\bm q}_{\phi(u)}$ is the semantic-level attention vector for node type $\phi(u)$. In comparison with HAN, MAGNN employs an additional projection to get the final representation $\bme_u$.

For link prediction, MAGNN adopts the loss introduced in GraphSAGE \cite{hamilton2017inductive}, which is equivalent to the objective in Eq.~(\ref{eqn:graphsage}).

\vspace{1mm}

\noindent \textbf{HGT \cite{hu2020heterogeneous}.} Inspired by the success of Transformer \cite{vaswani2017attention,devlin2019bert} in text representation learning, Hu et al. propose to use each edge's type to parameterize the Transformer-like self-attention architecture. To be specific, for each edge $(u, v)$, their Heterogeneous Graph Transformer (HGT) model maps $v$ into a \textit{Query} vector, and $u$ into a \textit{Key} vector, and calculate their dot product as attention:
\begin{equation}
\begin{gathered}
{\bm Q}_v^i = Q_{\phi(v)}^i (\bmh_v^{(k)}),\ \ \ {\bm K}_u^i = K_{\phi(u)}^i (\bmh_u^{(k)}) \\
{\rm Head}^{{\rm ATT}}_i(u,v) = \Big(\frac{{\bm K}_u^i\bmW^{{\rm ATT}}_{\psi(u,v)}{\bm Q}_v^{i\ T}}{\sqrt{d}}\Big)\mu(\phi(u),\psi(u,v), \phi(v)), \\
{\bf Attention}(u,v) = {\rm Softmax}_{u \in N(v)}\Big({\big|\big|}_i {\rm Head}^{{\rm ATT}}_i(u,v) \Big). \notag
\end{gathered}
\end{equation}
Here, $\bmh_u^{(k)}$ is the output of the $k$-th HGT layer ($\bmh_u^{(0)} = \bmx_u$); $Q_{\phi(v)}^i$ and $K_{\phi(u)}^i$ are node type-aware linear mappings; ${\rm Head}^{{\rm ATT}}_i$ is the $i$-th attention head; $\mu$ is a prior tensor representing the weight of each edge type in the attention. Parallel to the calculation of attention, the message passing process can be computed in a similar way by incorporating node and edge types:
\begin{equation}
\begin{gathered}
{\rm Head}^{{\rm MSG}}_i(u,v) = M_{\phi(u)}^i (\bmh_u^{(k)})\bmW^{{\rm MSG}}_{\psi(u,v)}, \\
{\bf Message}(u,v) = {\big|\big|}_i {\rm Head}^{{\rm MSG}}_i(u,v), \notag
\end{gathered}
\end{equation}
where $M_{\phi(u)}^i$ is also a node type-aware linear mapping. To aggregate the messages from $v$'s neighborhood, the attention vector serves as the weight to get the updated vector:
\begin{equation}
    \hat{\bmh}^{(k+1)}_v = \sum_{u\in N(v)} {\bf Attention}(u,v)\odot {\bf Message}(u,v), \notag
\end{equation}
and following the residual connection \cite{he2016deep}, the output of the $k$-th layer is
\begin{equation}
    \bmh^{(k+1)}_v = A_{\phi(v)}(\sigma(\hat{\bmh}^{(k+1)}_v))+ \bmh^{(k)}_v. \notag
\end{equation}
Here, $A_{\phi(u)}$ is a linear function mapping $v$'s vector back to its node type-specific distribution.

For the unsupervised link prediction task, HGT borrows the objective function from Neural Tensor Network \cite{socher2013reasoning}, which will be further discussed in Section \ref{sec:relation}.

\begin{table*}[h!]
	\centering
	\scalebox{0.925}{
	\begin{tabular}{c|c|c|c|c}
		\hline
		{\bf Algorithm} & $w_{uv}^{(l)}$ & $d(\bme_u,\bme_v)$ & $\mathcal{J}_{R0}$ & {\bf Applications}\\
		\hline
		TransE \cite{bordes2013translating}  & \multirow{6}{*}{$\textbf{1}_{(u,v)\in E_l}$}  &
		$||\bme_u+\bme_l-\bme_v||$ & $\sum_v(||\bme_v||-1)$ &
		\multirow{6}{*}{\begin{tabular}[c]{@{}c@{}} KB completion, \\ relation extraction \\ from text \end{tabular}}\\
		\cline{1-1} \cline{3-4}
		TransH \cite{wang2014knowledge} & & \makecell{$||\bme_{u,l}+\bme_l-\bme_{v,l}||^2$, \\ $\bme_{v,l} = \bme_v - \bmw_l^T\bme_v\bmw_l$} & \makecell{$\sum_l(||\bmw_l||-1) + \sum_v[||\bme_v||-1]_+$ \\ $+\sum_l\Big[\frac{(\bmw_l^T\bme_l)^2}{||\bme_l||^2}-\epsilon^2\Big]_+$}\\
		\cline{1-1} \cline{3-4}
		TransR \cite{lin2015learning} & & \makecell{$||\bme_{u,l}+\bme_l-\bme_{v,l}||^2$, \\$\bme_{v,l} = \bmA_l\bme_v$} & \makecell{$\sum_v[||\bme_v||-1]_+ + \sum_l[||\bme_l||-1]_+$ \\ $+\sum_v[||\bmA_l\bme_v||-1]_+$} \\
		\hline
		RHINE \cite{lu2019relation} & \makecell{edge weight of \\ $(u,v)$ with type $l$} & \makecell{$||\bme_u-\bme_v||^2$ if $l$ models affiliation, \\ $||\bme_u+\bme_l-\bme_v||$ if $l$ models interaction} & N/A & \makecell{link prediction, \\ node classification}\\
		\hline
		RotatE \cite{sun2019rotate} & \multirow{19}{*}{$\textbf{1}_{(u,v)\in E_l}$} & $||\bme_u\odot \bme_l - \bme_v||^2$, \ \  $\bme_u, \bme_v, \bme_l \in \mathbb{C}^n$ & $\sum_l(||\bme_l||-1)$ & \makecell{KB completion, \\ relation pattern \\ inference} \\
		\cline{1-1} \cline{3-5}
		RESCAL \cite{nickel2011three} &  & $||\sqrt{\bmA_l}(\bme_u - \bme_v)||^2$ & $\sum_v[||\bme_v||-1]_+ + \sum_l [||\bmA_l||_F -1]_+ $ & \makecell{entity resolution, \\ link prediction} \\
		\cline{1-1} \cline{3-5}
		DistMult \cite{yang2014embedding} & {\color{white}\Big(} & $||\sqrt{\bmA_l}(\bme_u - \bme_v)||^2$, \ \ $\bmA_l = diag(\bme_l)$ & $\sum_v(||\bme_v||-1) + \sum_l [||\bme_l|| -1]_+$ & \multirow{15}{*}{\begin{tabular}[c]{@{}c@{}} KB completion, \\ triplet classification \end{tabular}} \\
		\cline{1-1} \cline{3-4}
		HolE \cite{nickel2016holographic} & {\color{white}\Big(}  & $||\bme_r-\mathcal{F}^{-1}(\overline{\mathcal{F}(\bme_u)}\odot \mathcal{F}(\bme_v))||^2$ & $\sum_v[||\bme_v||-1]_+ + \sum_l [||\bme_l|| -1]_+$ &  \\
		\cline{1-1} \cline{3-4}
		ComplEx \cite{trouillon2016complex} & {\color{white}\Big(} & $||\bmA_l\bme_u - \bme_v||^2$, \ \ $\bmA_l = diag(\bme_l)$, \ \ $\bme_u, \bme_v, \bme_l \in \mathbb{C}^n$ & \multirow{4}{*}{$\sum_v||\bme_v||^2 + \sum_l ||\bme_l||^2$} &  \\
		\cline{1-1} \cline{3-3}
		SimplE \cite{kazemi2018simple} &  & \makecell{$\frac{1}{2}\Big(||\sqrt{\bmA_l}(\bme_u - \bme_v)||^2 + ||\sqrt{\bmA_{l^{-1}}}(\bme_v - \bme_u)||^2 \Big)$, \\ $\bmA_l = diag(\bme_l)$, \ \ $\bmA_{l^{-1}} = diag(\bme_{l^{-1}})$} &  &  \\
		\cline{1-1} \cline{3-4}
		TuckER \cite{balazevic2019tucker} & {\color{white}\Big(} & $C-\mathcal{W} \times_1 \bme_u \times_2 \bme_l \times_3 \bme_v$ & N/A &  \\
		\cline{1-1} \cline{3-4}
		NTN \cite{socher2013reasoning} &  & $C- \bme_l^T$tanh$\Big(\bme_u^T\mathcal{M}_l\bme_v+{\bm M}_{l,1}\bme_u+{\bm M}_{l,2}\bme_v+\bmb_l\Big)$ & 
		$||\Theta||_2^2$ &  \\
		\cline{1-1} \cline{3-4}
		ConvE \cite{dettmers2018convolutional} &  & $C-\sigma\Big({\rm vec}\big(\sigma([{\bm E}_u; {\bm E}_l] * \omega)\big)\bmW\Big)\bme_v$ & N/A & \\
		\cline{1-1} \cline{3-4}
		NKGE \cite{wang2018knowledge} & & \makecell{same as TransE or ConvE, where \\ $\bme_u = \sigma({\bm g}_u) \odot \bme_u^s + \big(1-\sigma({\bm g}_u)\big) \odot \bme_u^n$} & $||\Theta||_2^2$ &   \\
		\cline{1-1} \cline{3-4}
		SACN \cite{shang2019end} & & $C-f\Big({\rm vec}\big({\bm M}(\bme_u, \bme_l)\big)\bmW\Big)\bme_v$ & N/A & \\
		\hline
	\end{tabular}
    }
	\caption{\label{tab:translation} A summary of relation-learning based HNE algorithms. \textnormal{(Additional notations: $f$: a non-linear activation function; $[x]_+$: $\max\{x, 0\}$; $||\cdot||_F$: the Frobenius norm of a matrix; $\mathcal{F}, \mathcal{F}^{-1}$: the fast Fourier transform and its inverse; $\overline{x}$: the complex conjugate of $x$; $l^{-1}$: the inverse of relation $l$; $\times_i$: the tensor product along the $i$-th mode; $\Theta$: the set of learned parameters; ${\bm E}_u$, ${\bm E}_l$: 2D reshaping matrices of $\bme_u$ and $\bme_l$ \protect\cite{dettmers2018convolutional}; $\rm vec$: the vectorization operator; $*$: the convolution operator; ${\bm M}(\bme_u, \bme_l)$: a matrix aligning the output vectors from the convolution with all kernels \protect\cite{shang2019end}.)}}
\end{table*}

Some other message-passing approaches are summarized in Table \ref{tab:message}. For example, HEP \cite{zheng2018heterogeneous} aggregates $u$'s representation from $\mathcal{N}_o(u)$ (\ie, the \textit{node type-aware} neighborhood) to reconstruct $u$'s own embedding; HetGNN \cite{zhang2019heterogeneous} also adopts a node type-aware neighborhood aggregation, but the neighborhood is defined through random walk with restart \cite{tong2006fast}; GATNE \cite{cen2019representation} aggregates $u$'s representation from $\mathcal{N}_l(u)$ (\ie, the \textit{edge type-aware} neighborhood) and is applicable to both transductive and inductive network embedding settings; 
MV-ACM \cite{zhao2020deep} aggregates $u$'s representation from $\mathcal{N}_{\mathcal{M}}(u)$ (\ie, the \textit{meta-path-aware} neighborhood) and utilizes an adversarial learning framework to learn the reciprocity between different edge types.

The objective of message-passing approaches mentioned above can also be written as Eq.~(\ref{eqn:obj}) (except HGT, whose objective is the same as NTN \cite{socher2013reasoning} and will be discussed in Section \ref{sec:relation}), where $s(u,v)$ is a function of $\bme_u$ and $\bme_v$. The only difference is that $\bme_u$ here is aggregated from $\bmx_v$ using GNNs. Following the derivation of proximity-preserving approaches, if we still write $\mathcal{J}_{R}$ in Eq.~(\ref{eq:general}) as the sum of $-\mathcal{J}_{R_0}$, $-\mathcal{J}_{R_1}$ and $\mathcal{J}_{R_2}$, we can get the exactly same $\mathcal{J}_{R_1}$ and $\mathcal{J}_{R_2}$ as in Eq.~(\ref{eq:ppj}). We summarize the choices of $w_{uv}$, $d(\bme_u,\bme_v)$ and the aggregation function in Table \ref{tab:message}.

Within this group of algorithms, HEP has an additional reconstruction loss $\mathcal{J}_{R_0}=\sum_v ||\tilde{\bme}_v-\bme_v||^2$, and MV-ACM \cite{zhao2020deep} has an adversarial loss $\mathcal{J}_{R_0}=\sum_{v \in V, l \in \mathcal{T}_E} \mathbb{E}_{Z\sim \mathcal{G}(\cdot|l, v)}$ $\big[\log(1-\mathcal{D}(Z,l,v))\big]$, where $\mathcal{G}$ and $\mathcal{D}$ are the generator and discriminator of complementary edge types, respectively. All the other algorithms in this group have $\mathcal{J}_{R_0}=0$.

Similar to the case of proximity-preserving approaches, there are also message-passing methods adopting other forms of objectives. For example, GraphInception \cite{zhang2018deep} studies collective classification (where the labels within a group of instances are correlated and should be inferred collectively) using meta-path-aware aggregation; HDGI \cite{ren2019heterogeneous} maximizes local-global mutual information to improve unsupervised training based on HAN; NEP \cite{yang2019neural} performs semi-supervised node classification using an edge type-aware propagation function to mimic the process of label propagation; NLAH \cite{xiao2019non} also considers semi-supervised node classification by extending HAN with several pre-computed non-local features; HGCN \cite{zhu2020hgcn} explores the collective classification task and improves GraphInception by considering semantics of different types of edges/nodes and relations among different node types; HetETA \cite{hong2020heteta} studies a specific task of estimating the time of arrival in intelligent transportation using a heterogeneous graph network with fast localized spectral filtering.

\subsection{Relation-Learning Methods}
\label{sec:relation}

As discussed above, knowledge graphs can be regarded as a special case of heterogeneous networks, which are schema-rich \cite{wang2016relsim}. To model network heterogeneity, existing knowledge graph embedding approaches explicitly model the relation types of edges via parametric algebraic operators, which are often shallow network embedding \cite{wang2017knowledge,ji2020survey}. Compared to the shallow proximity-preserving HNE models, they often focus on the designs of triplet based scoring functions instead of meta-paths or meta-graphs due to the large numbers of entity and relation types.

Each edge in a heterogeneous network can be viewed as a triplet $(u,l,v)$ composed of two nodes $u,v \in V$ and an edge type $l \in \mathcal{T}_E$ (\ie, entities and relations, in the terminology of KG). The goal of relation-learning methods is to learn a scoring function $s_l(u,v)$ which evaluates an arbitrary triplet and outputs a scalar to measure the acceptability of this triplet. This idea is widely adopted in KB embedding. Since there are surveys of KB embedding algorithms already \cite{wang2017knowledge}, we only cover the most popular approaches here and highlight their connections to HNE.
\vspace{1mm}

\noindent \textbf{TransE \cite{bordes2013translating}.} TransE assumes that the relation induced by $l$-labeled edges corresponds to a translation of the embedding (\ie, $\bme_u+\bme_l \approx \bme_v$) when $(u,l,v)$ holds. Therefore, the scoring function of TransE is defined as
\begin{equation}
s_l(u,v)=-||\bme_u+\bme_l-\bme_v||_p, 
\end{equation}
where $p=1$ or $2$. The objective is to minimize a margin based ranking loss.
\begin{equation}
\mathcal{J} = \sum_{(u,l,v) \in T}\sum_{(u',l,v') \in T'_{(u,l,v)}} \max(0, \gamma-s_l(u,v)+s_l(u',v')),
\label{eqn:margin}
\end{equation}
where $T$ is the set of positive triplets (\ie, edges); $T'_{(u,l,v)}$ is the set of corrupted triplets, which are constructed by replacing either $u$ or $v$ with an arbitrary node. Formally,
\begin{equation}
T'_{(u,l,v)} = \{(u',l,v)|u'\in V\} \cup \{(u,l,v')|v'\in V\}. \notag
\end{equation}

TransE is the most representative model using ``a translational distance'' to define the scoring function. It has many extensions. For example, TransH \cite{wang2014knowledge} projects each entity vector to a relation-specific hyperplane when calculating the distance; TransR \cite{lin2015learning} further extends relation-specific hyperplanes to relation-specific spaces; RHINE \cite{lu2019relation} distinguishes affiliation relations from interaction relations and adopts different objectives for the two types of relations. For more extensions of TransE, please refer to \cite{wang2018knowledge}. 

Recently, Sun et al. \cite{sun2019rotate} propose the RotatE model, which defines each relation as a rotation (instead of a translation) from the source entity to the target entity in the complex vector space. Their model is able to describe various relation patterns including symmetry/antisymmetry, inversion, and composition.

\vspace{1mm}

\noindent \textbf{DistMult \cite{yang2014embedding}.} In contrast to translational distance models \cite{bordes2013translating,wang2014knowledge,lin2015learning}, DistMult exploits a similarity based scoring function. Each relation is represented by a diagonal matrix $\bmA_l=diag(\bme_{l1},...,\bme_{ld})$, and $s_l(u,v)$ is defined using a bilinear function:
\begin{equation}
s_l(u,v) = \bme_u^T\bmA_l\bme_v. \notag
\end{equation}
Note that $s_l(u,v) = s_l(v,u)$ for any $u$ and $v$. Therefore, DistMult is mainly designed for symmetric relations.

Using the equation $||\bmx-\bmy||^2 = (\bmx-\bmy)^T(\bmx-\bmy) = ||\bmx||^2+||\bmy||^2-2\bmx^T\bmy$, we have
\begin{equation}
\begin{split}
  s_l(u,v) &= (\sqrt{\bmA_l}\bme_u)^T(\sqrt{\bmA_l}\bme_v) \\
  &= \frac{1}{2}\big(||\sqrt{\bmA_l}\bme_u||^2 + ||\sqrt{\bmA_l}\bme_v||^2 - ||\sqrt{\bmA_l}(\bme_u - \bme_v)||^2\big). \notag  
\end{split}
\end{equation}

\vspace{1mm}

\noindent \textbf{ComplEx \cite{trouillon2016complex}.} Instead of considering a real-valued embedding space, ComplEx introduces complex-valued representations for $\bme_u$, $\bme_v$ and $\bme_l$. Similar to DistMult, it utilizes a similarity based scoring function:
\begin{equation}
s_l(u,v) = {\rm Re}(\bme_u^T\bmA_l\overline{\bme_v}), \notag
\end{equation}
where ${\rm Re}(\cdot)$ is the real part of a complex number; $\overline{\bme}$ is the complex conjugate of $\bme$; $\bmA_l=diag(\bme_{l1},...,\bme_{ld})$. Here, it is possible that $s_l(u,v) \neq s_l(v,u)$, which allows ComplEx to capture asymmetric relations.

In the complex space, using the equation $||\bmx-\bmy||^2 = (\bmx-\bmy)^T(\overline{\bmx-\bmy}) =  ||\bmx||^2+||\bmy||^2-\bmx^T\overline{\bmy}-\bmy^T\overline{\bmx} = ||\bmx||^2+||\bmy||^2-2{\rm Re}(\bmx^T\overline{\bmy})$, we have
\begin{equation}
\begin{split}
  s_l(u,v) &={\rm Re}((\bmA_l\bme_u)^T\overline{\bme_v}) \\
  &= \frac{1}{2}\big(||\bmA_l\bme_u||^2 + ||\bme_v||^2 - ||\bmA_l\bme_u - \bme_v||^2\big). \notag  
\end{split}
\end{equation}

Besides ComplEx, there are many other extensions of DistMult. RESCAL \cite{nickel2011three} uses a bilinear scoring function similar to the one in DistMult, but $\bmA_l$ is no longer restricted to be diagonal; HolE \cite{nickel2016holographic} composes the node representations using the circular
correlation operation, which combines the expressive power of RESCAL with the efficiency of DistMult; SimplE \cite{kazemi2018simple} considers the inverse of relations and calculates the average score of $(u,l,v)$ and $(v, l^{-1}, u)$; TuckER \cite{balazevic2019tucker} proposes to use a three-way Tucker tensor decomposition approach to learning node and relation embeddings.

\vspace{1mm}

\noindent \textbf{ConvE \cite{dettmers2018convolutional}.} ConvE goes beyond simple distance or similarity functions and proposes deep neural models to score a triplet. The score is defined by a convolution over 2D shaped embeddings. Formally, 
\begin{equation}
s_l(u,v) = \sigma({\rm vec}(\sigma([{\bm E}_u; {\bm E}_r] * \omega))\bmW)\bme_v, \notag
\end{equation}
where ${\bm E}_u$ and ${\bm E}_r$ denote the 2D reshaping matrices of node embedding and relation embedding, respectively; $\rm vec$ is the vectorization operator that maps a $m$ by $n$ matrix to a $mn$-dimensional vector; ``$*$'' is the convolution operator.

There are several other models leveraging deep neural scoring functions. For example, NTN \cite{socher2013reasoning} proposes to combine the two node embedding vectors by a relation-specific tensor $\mathcal{M}_l \in \mathbb{R}^{d\times d\times d_l}$, where $d_l$ is the dimension of $\bme_l$; NKGE \cite{wang2018knowledge} develops a deep memory network to encode information from neighbors and employs a gating mechanism to integrate structure representation $\bme_u^s$ and neighbor representation $\bme_u^n$; SACN \cite{shang2019end} encodes node representations using a graph neural network and then scores the triplet using a convolutional neural network with the translational property.

Most relation-learning approaches adopt a margin based ranking loss with some regularization terms that generalizes Eq.~(\ref{eqn:margin}):
\begin{equation}
\sum_{(u,l,v)}w_{uv}^{(l)} \sum_{(u',l,v')} \max(0, \gamma-s_l(u,v)+s_l(u',v')) + \mathcal{J}_{R0}.
\label{eqn:robj}
\end{equation}
In \cite{qiu2018revisiting}, Qiu et al. point out that the margin based loss shares a very similar form with the following negative sampling loss:
\begin{equation}
- \sum_{(u,l,v)} \Big(\log(\sigma(s_l(u,v))) - b\mathbb{E}_{(u',l,v')}\big[\log(\sigma(s_l(u',v')))\big]\Big). \notag
\end{equation}
Following \cite{qiu2018revisiting}, if we use the negative sampling loss to rewrite Eq.~(\ref{eqn:robj}), we are approximately maximizing
\begin{equation}
\mathcal{J} = \sum_{(u,l,v)} w_{uv}^{(l)} \log \frac{\exp(s_l(u,v))}{\sum_{(u',l,v')} \exp(s_l(u',v'))} + \mathcal{J}_{R0}. \notag
\end{equation}

For translational models \cite{bordes2013translating,wang2014knowledge,lin2015learning,lu2019relation} whose $s_l(u,v)$ is described by a distance function, maximizing $\mathcal{J}$ is equivalent to
\begin{equation}
\begin{split}
\min -\mathcal{J} &= \sum_{(u,l,v)} w_{uv}^{(l)} \underbrace{||\bme_u+\bme_l-\bme_v||^{1\text{ or }2}}_{d(\bme_u,\bme_v)} - \mathcal{J}_{R0} \\
&+ \underbrace{\sum_{(u,l,v)} w_{uv}^{(l)}\log\sum_{(u',l,v')} \exp(s_l(u',v'))}_{\mathcal{J}_{R1}}.
\end{split}
\label{eqn:tpj}
\end{equation}
In this case, we can write $\mathcal{J}_R$ as $-\mathcal{J}_{R0}+\mathcal{J}_{R1}$. For RotatE \cite{sun2019rotate}, the objective is the same except that $d(\bme_u,\bme_v) = ||\bme_u\odot \bme_l-\bme_v||^2$.

For similarity based approaches \cite{nickel2011three,yang2014embedding,trouillon2016complex,kazemi2018simple}, we follow the derivation of Eq.~(\ref{eq:ppj}), and the objective will be
\begin{equation}
\begin{split}
\min -\mathcal{J} &= \sum_{(u,l,v)} \frac{w^{(l)}_{uv}}{2}\underbrace{||f(\bme_u) - f(\bme_v)||^2}_{d(\bme_u,\bme_v)} - \mathcal{J}_{R0} \\
& - \underbrace{\sum_{{(u,l,v)}} \frac{w^{(l)}_{uv}}{2} \Big(||f(\bme_u)||^2 + ||f(\bme_v)||^2\Big)}_{\mathcal{J}_{R1}} \\
& + \underbrace{\sum_{{(u,l,v)}} w^{(l)}_{uv} \log \sum_{{(u',l,v')}} \exp(s_l(u',v'))}_{\mathcal{J}_{R2}}. \notag
\end{split}
\end{equation}
Here, $f(\bme_u) = \sqrt{\bmA_l}\bme_u$ in RESCAL \cite{nickel2011three} and DistMult \cite{yang2014embedding}; $f(\bme_u) = \sqrt{\bmA_l}\bme_u$ or $\sqrt{\bmA_{l^{-1}}}\bme_u$ in SimplE \cite{kazemi2018simple}; $f(\bme_u) = \bmA_l\bme_u$ and $f(\bme_v) = \bme_v$ in ComplEx \cite{trouillon2016complex}. The regularization term is $\mathcal{J}_R = -\mathcal{J}_{R0}-\mathcal{J}_{R1}+\mathcal{J}_{R2}$.

For neural triplet scorers \cite{dettmers2018convolutional,socher2013reasoning,oh2018knowledge,shang2019end}, the forms of $s_l(u,v)$ are more complicated than translational distances or bilinear products. In these cases, since distance (or dissimilarity) and proximity can be viewed as reverse metrics, we define $d(\bme_u, \bme_v)$ as $C-s_l(u,v)$, where $C$ is an constant upper bound of $s_l(\cdot, \cdot)$. 
Then the derivation of the loss function is similar to that of Eq.~(\ref{eqn:tpj}), \ie, 
\begin{equation}
\begin{split}
\min -\mathcal{J} &= \sum_{(u,l,v)} w_{uv}^{(l)} \underbrace{\Big(C-s_l(u,v)\Big)}_{d(\bme_u,\bme_v)} - \mathcal{J}_{R0} \\
&+ \underbrace{\sum_{(u,l,v)} w_{uv}^{(l)}\log\sum_{(u',l,v')} \exp(s_l(u',v'))}_{\mathcal{J}_{R1}}. \notag
\end{split}
\end{equation}
In this case, $\mathcal{J}_R= -\mathcal{J}_{R0}+\mathcal{J}_{R1}$.

We summarize the choices of $w_{uv}^{(l)}$, $d(\bme_u,\bme_v)$ and $\mathcal{J}_{R0}$ in Table \ref{tab:translation}. Note that for relation learning methods, $d(\cdot, \cdot)$ is usually \textit{not} a distance metric. For example, $d(\bme_u,\bme_v) \neq d(\bme_v,\bme_u)$ in most translational distance models and deep neural models. This is intuitive because $(u,l,v)$ and $(v,l,u)$ often express different meanings.
\section {Benchmark} 
\label{sec:data}
\subsection{Dataset Preparation}
Towards off-the-shelf evaluation of HNE algorithms with standard settings, in this work, we collect, process, analyze, and publish four new real-world heterogeneous network datasets from different domains, which we aim to set up as a handy and fair benchmark for existing and future HNE algorithms.\textsuperscript{\ref{repo}}

\begin{table*}[h!]
\centering
\begin{tabular}{l|cccccccc}
\hline
{\bf Dataset} & \#node type & \#node & \#link type& \#link & \#attributes & \#attributed nodes & \#label type & \#labeled node \\
\hline
DBLP & 4 & 1,989,077 & 6 & 275,940,913 & 300 & ALL & 13 & 618\\
\hline
Yelp & 4 & 82,465 & 4 & 30,542,675 & N/A & N/A & 16 & 7,417\\
\hline
Freebase & 8 & 12,164,758 & 36 & 62,982,566 & N/A & N/A & 8 & 47,190\\
\hline
PubMed & 4 & 63,109 & 10 & 244,986 & 200 & ALL & 8 & 454\\
\hline
\end{tabular}
\caption{\label{tab:stats}\textbf{A summary of the statistics on four real-world heterogeneous network datasets.}}
\end{table*}

\begin{figure*}[t!]
\centering
\subfigure[DBLP]{
\includegraphics[width=0.235\textwidth]{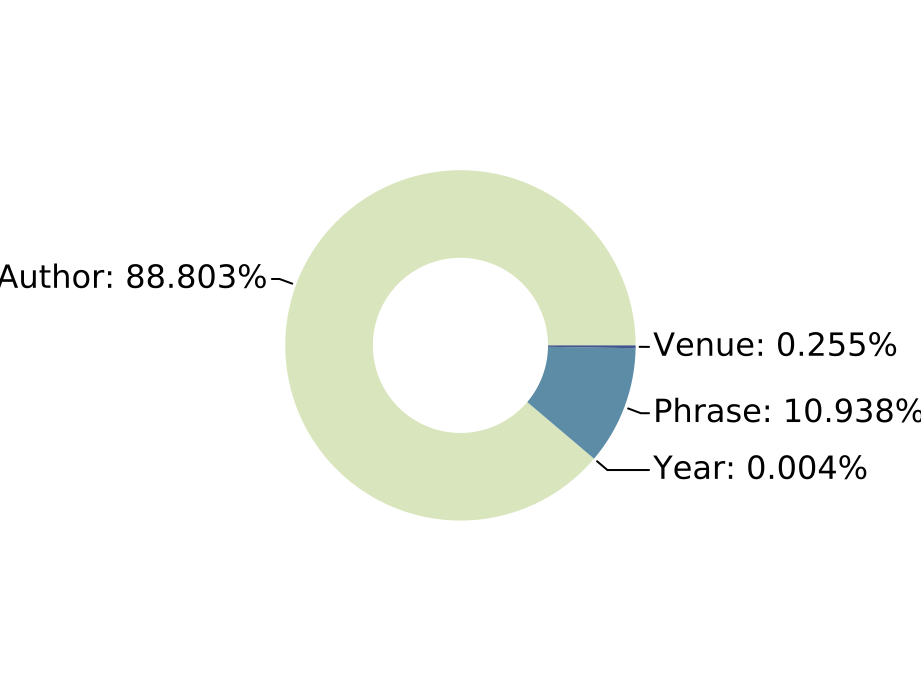}}
\subfigure[Yelp]{
\includegraphics[width=0.235\textwidth]{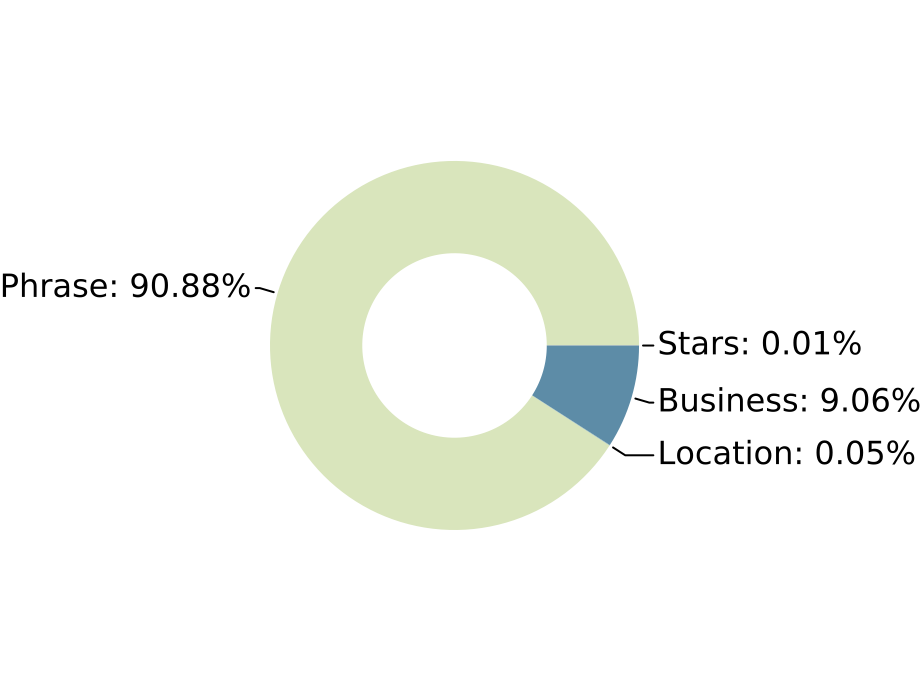}}
\subfigure[Freebase]{
\includegraphics[width=0.235\textwidth]{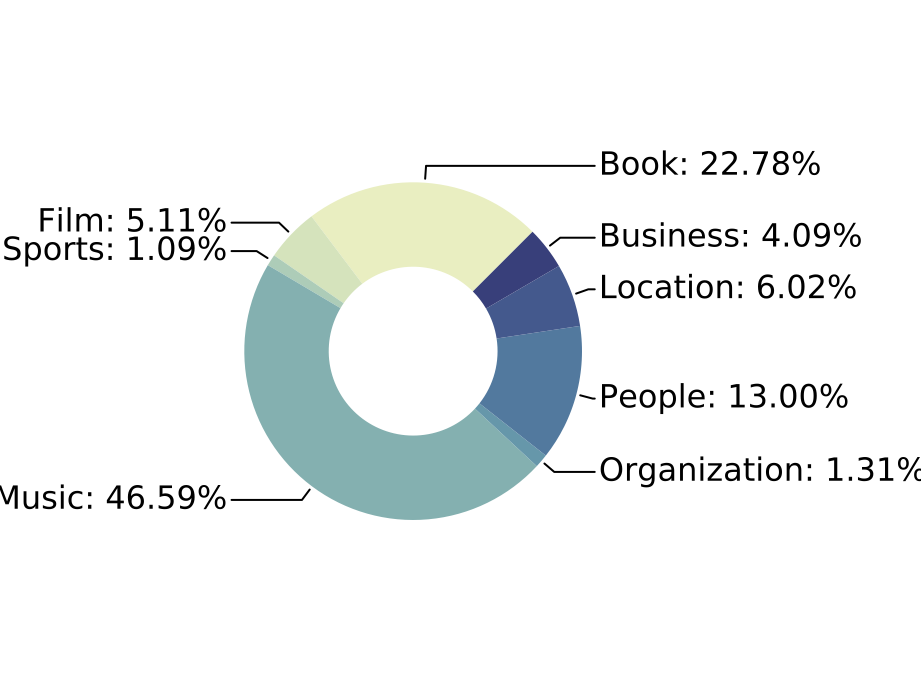}}
\subfigure[PubMed]{
\includegraphics[width=0.235\textwidth]{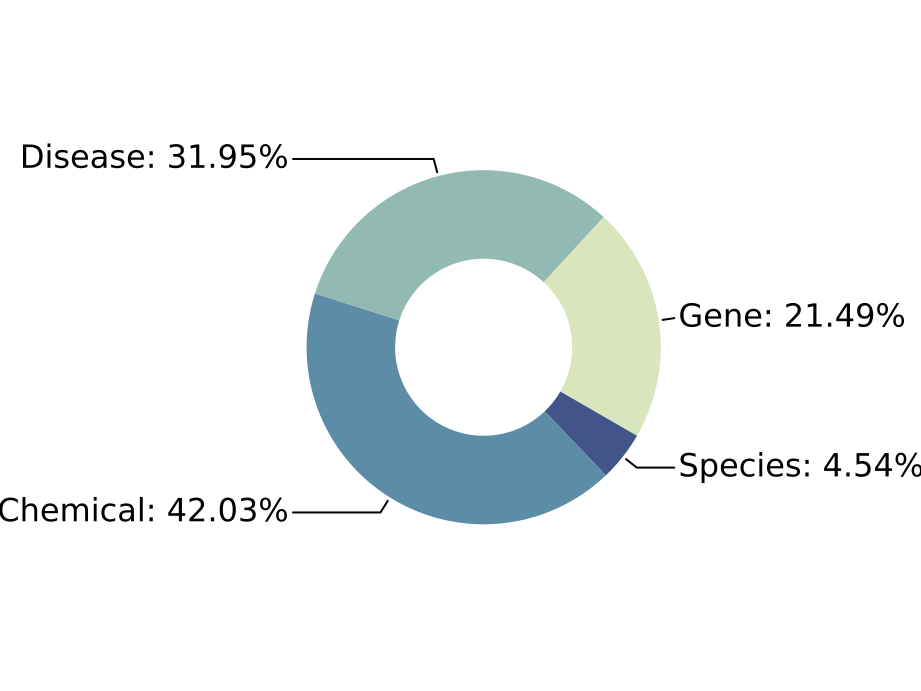}}
\caption{\textbf{Portion of different node types in four real-world heterogeneous network datasets.}}
\label{fig:nodetype}
\end{figure*}

\begin{figure*}[t!]
\centering
\vspace{10pt}
\subfigure[DBLP]{
\includegraphics[width=0.235\textwidth]{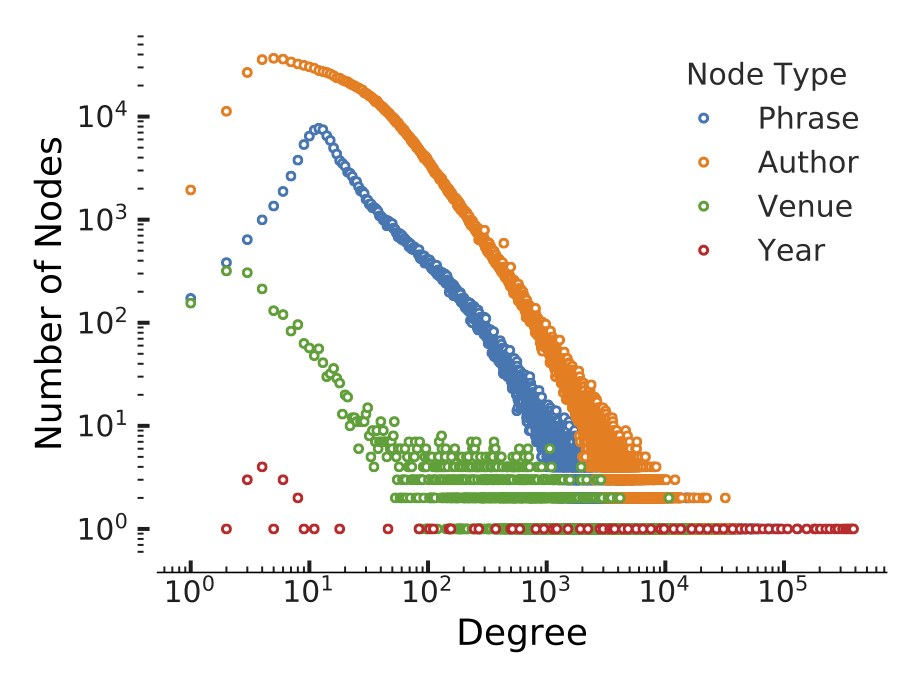}}
\subfigure[Yelp]{
\includegraphics[width=0.235\textwidth]{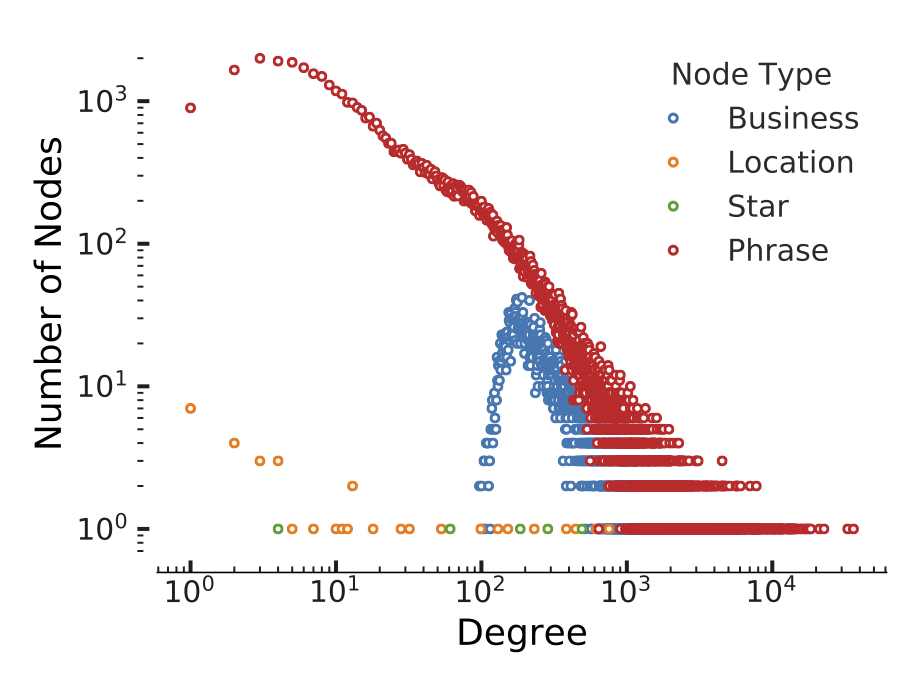}}
\subfigure[Freebase]{
\includegraphics[width=0.235\textwidth]{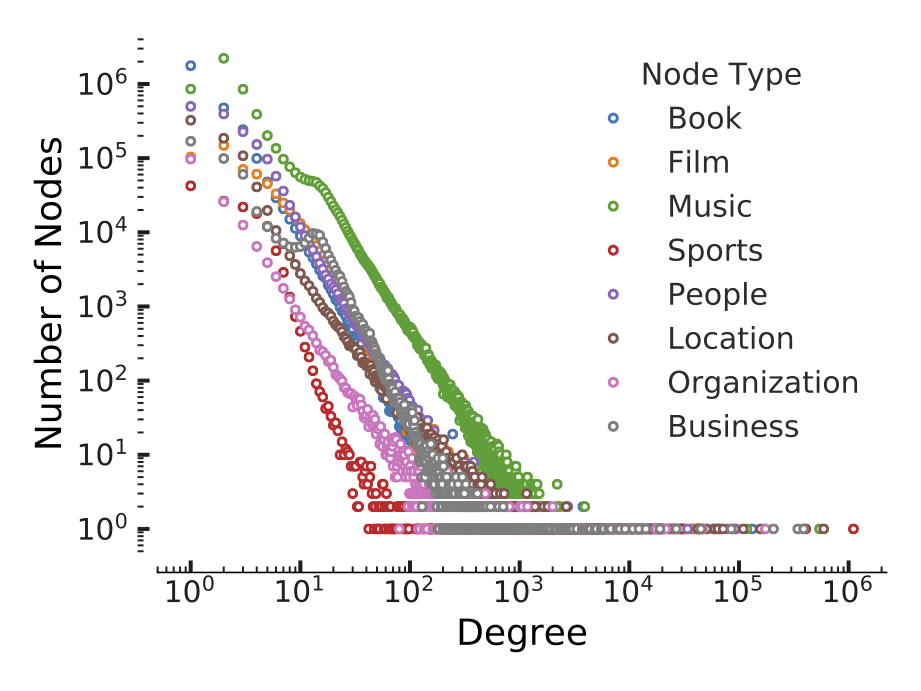}}
\subfigure[PubMed]{
\includegraphics[width=0.235\textwidth]{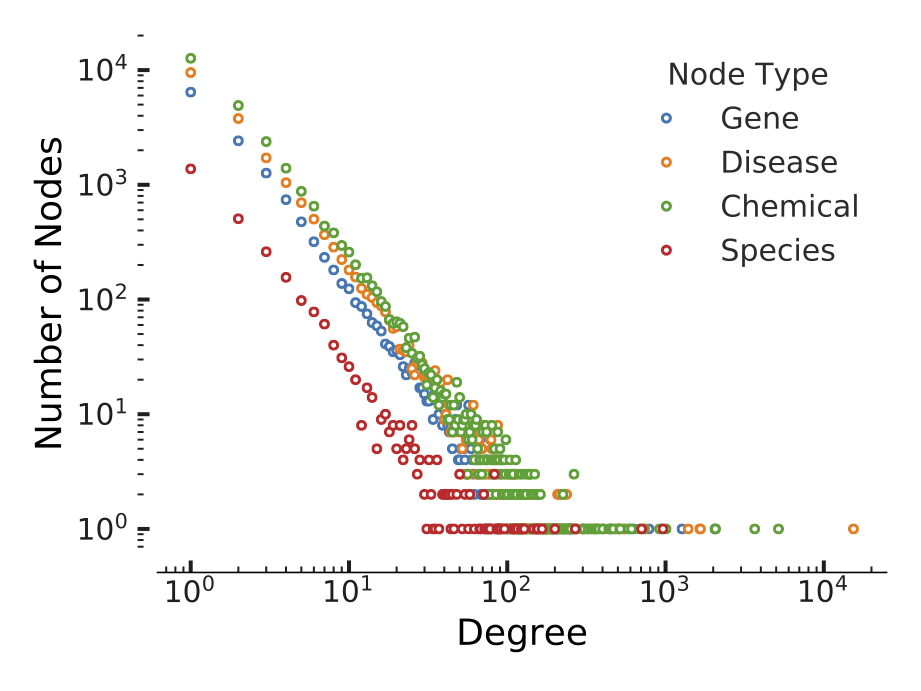}}
\caption{\textbf{Degree distribution of different node types in four real-world heterogeneous network datasets.}}
\label{fig:degree}
\end{figure*}

\begin{figure*}[t!]
\centering
\vspace{10pt}
\subfigure[DBLP]{
\includegraphics[width=0.235\textwidth]{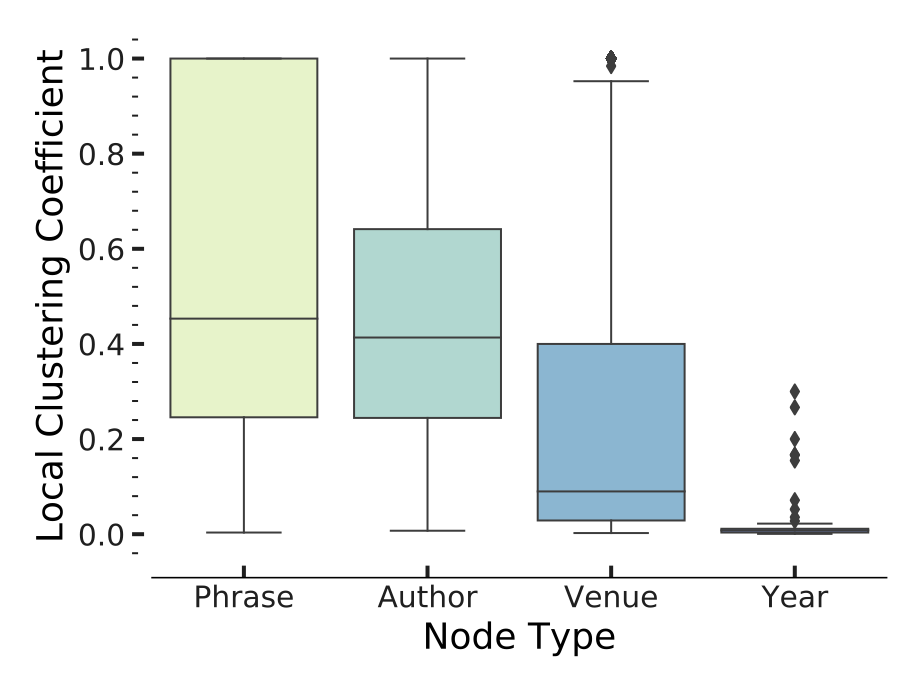}}
\subfigure[Yelp]{
\includegraphics[width=0.235\textwidth]{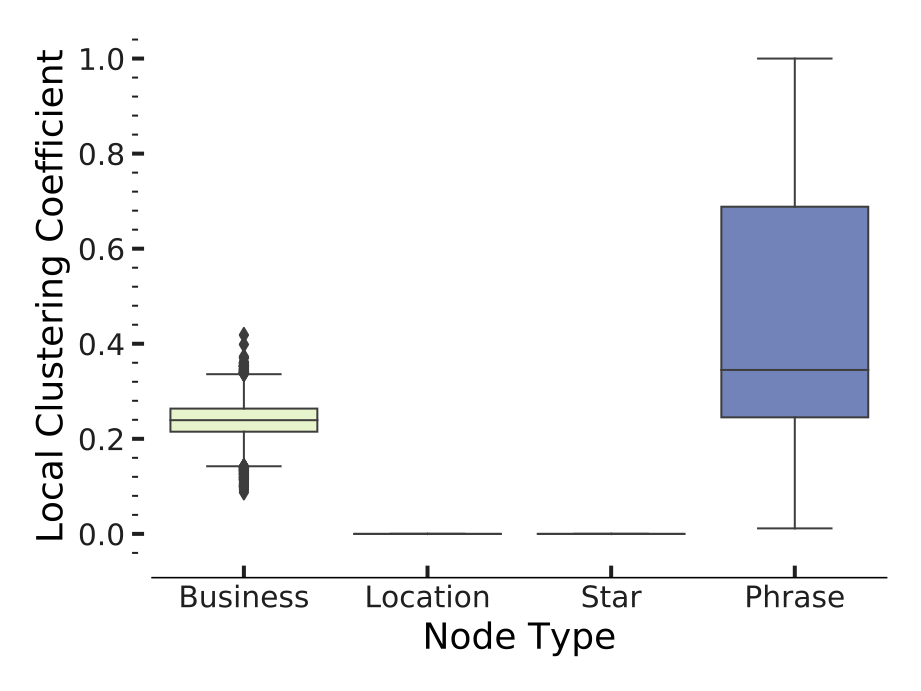}}
\subfigure[Freebase]{
\includegraphics[width=0.235\textwidth]{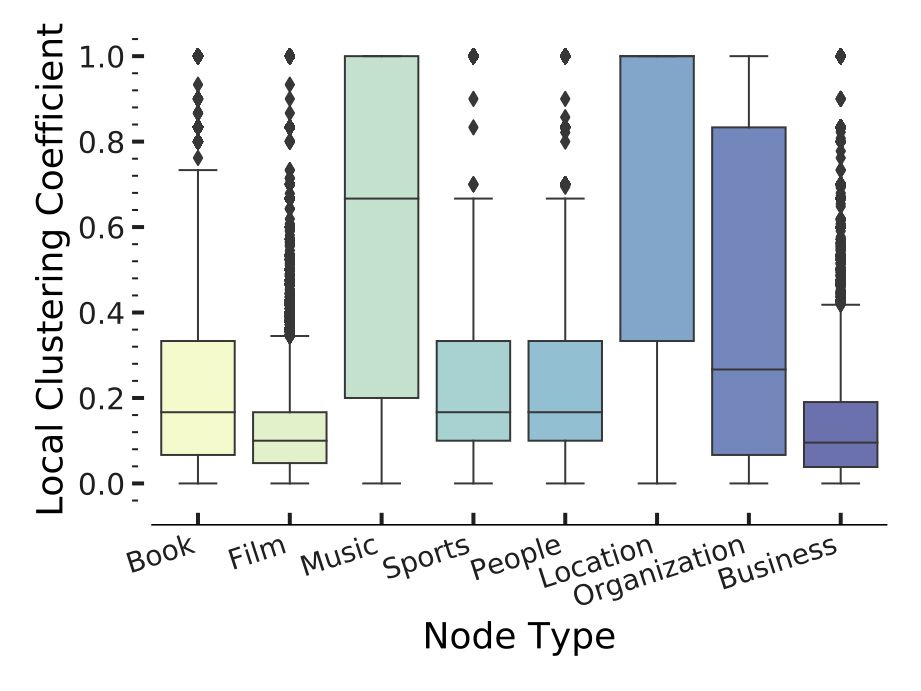}}
\subfigure[PubMed]{
\includegraphics[width=0.235\textwidth]{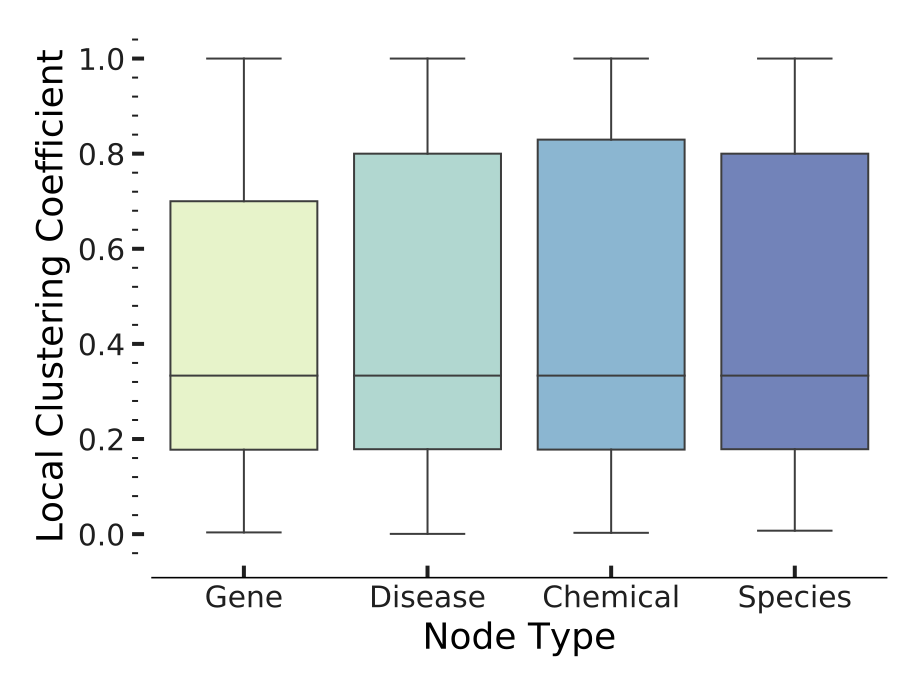}}
\caption{\textbf{Local clustering coefficient of different node types in four real-world heterogeneous network datasets.}}
\label{fig:cc}
\end{figure*}

\begin{figure*}[t!]
\centering
\vspace{10pt}
\subfigure[DBLP]{
\includegraphics[width=0.235\textwidth]{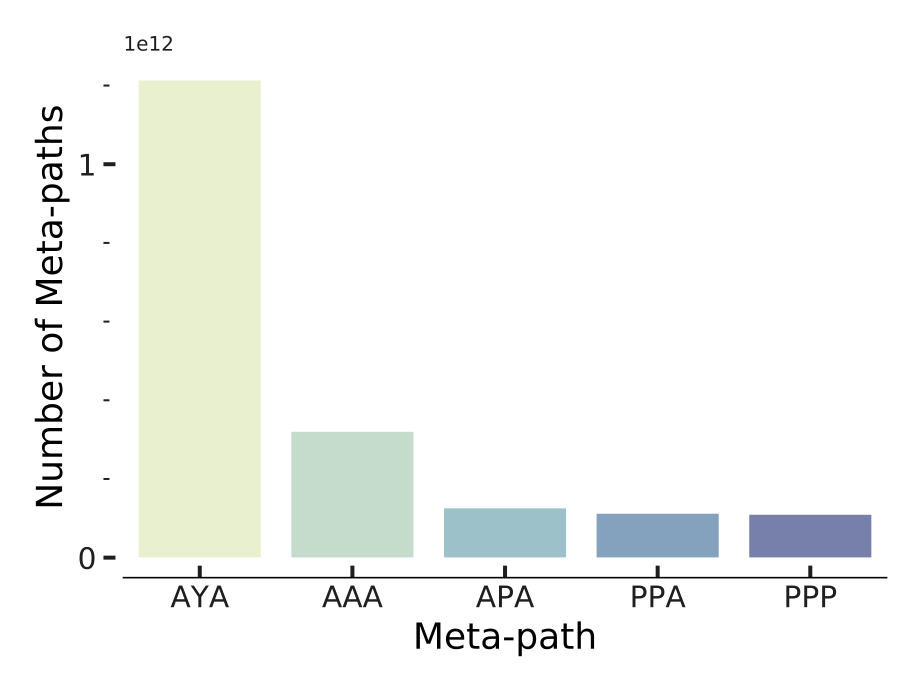}}
\subfigure[Yelp]{
\includegraphics[width=0.235\textwidth]{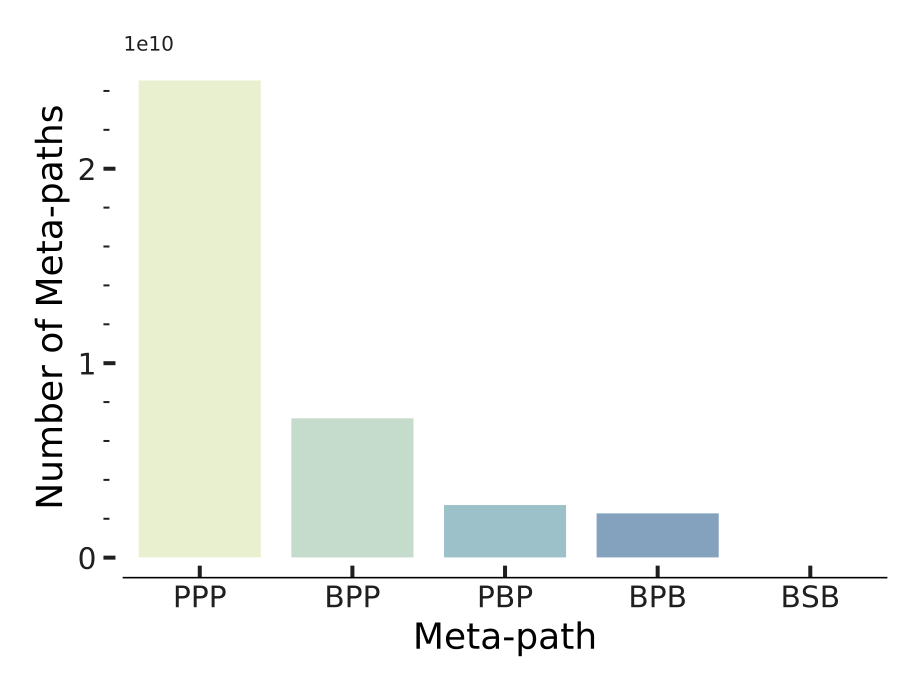}}
\subfigure[Freebase]{
\includegraphics[width=0.235\textwidth]{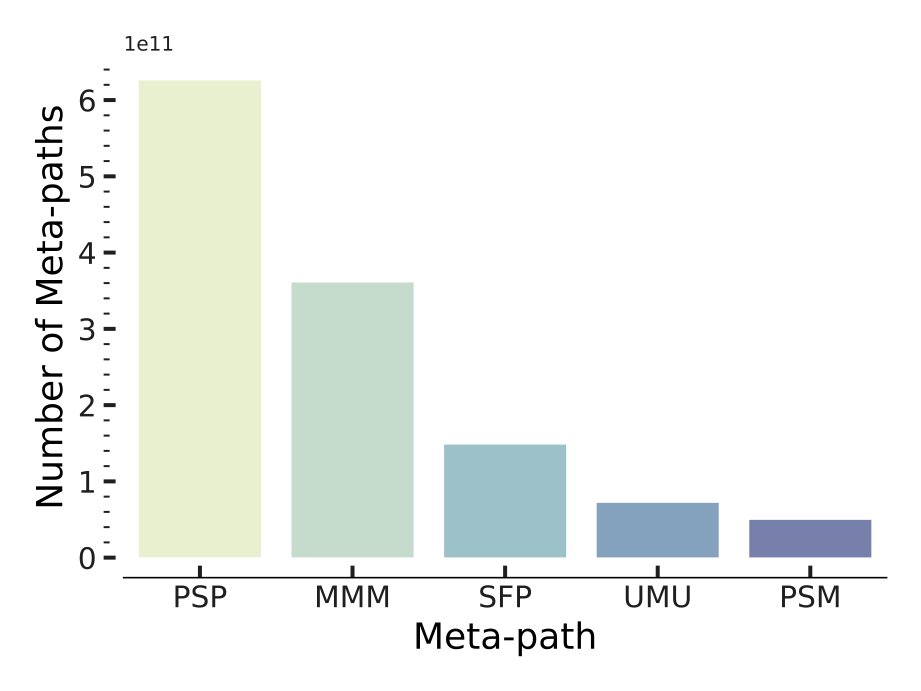}}
\subfigure[PubMed]{
\includegraphics[width=0.235\textwidth]{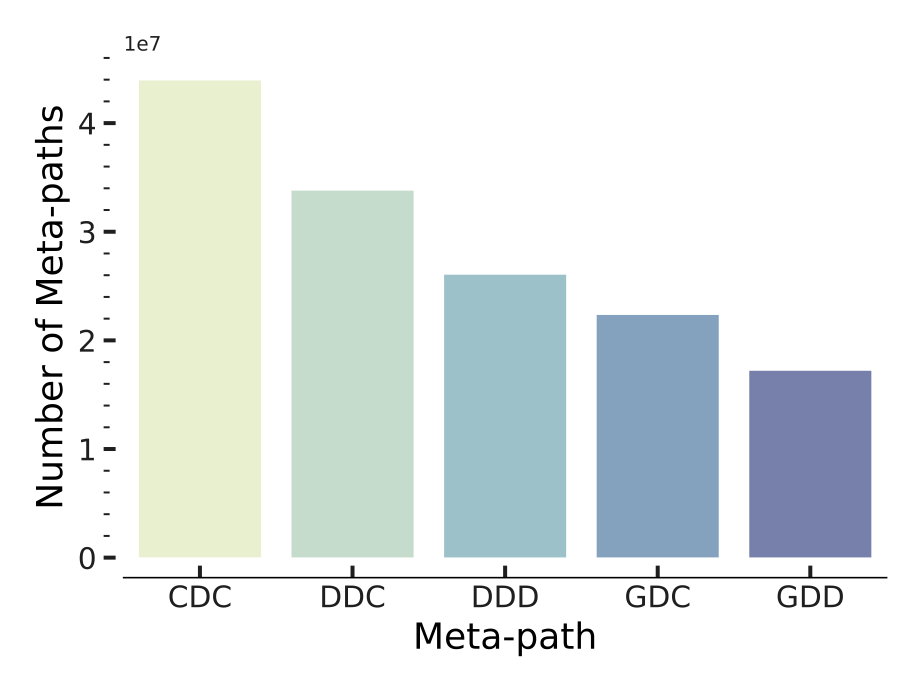}}
\caption{\textbf{Number of five most frequent 2-hop meta-paths in four real-world heterogeneous network datasets.}}
\label{fig:metapath}
\end{figure*}




\vspace{-3pt}
\header{DBLP.}
We construct a network of authors, papers, venues, and phrases from DBLP. Phrases are extracted by the popular AutoPhrase \cite{shang2017automated} algorithm from paper texts and further filtered by human experts. We compute word2vec \cite{mikolov2013distributed} on all paper texts and aggregate the word embeddings to get 300-dim paper and phrase features. Author and venue features are the aggregations of their corresponding paper features. We further manually label a relatively small portion of authors into 12 research groups from four research areas by crawling the web. Each labeled author has only one label.

\vspace{-3pt}
\header{Yelp.}
We construct a network of businesses, users, locations, and reviews from Yelp.\footnote{https://www.yelp.com/dataset/challenge} Nodes do not have features, but a large portion of businesses are labeled into sixteen categories. Each labeled business has one or multiple labels.

\vspace{-3pt}
\header{Freebase.}
We construct a network of books, films, music, sports, people, locations, organizations, and businesses from Freebase.\footnote{http://www.freebase.com/} Nodes are not associated with any features, but a large portion of books are labeled into eight genres of literature. Each labeled book has only one label.

\vspace{-3pt}
\header{PubMed.}
We construct a network of genes, diseases, chemicals, and species from PubMed.\footnote{https://www.ncbi.nlm.nih.gov/pubmed/} All nodes are extracted by AutoPhrase \cite{shang2017automated}, typed by bioNER \cite{wang2019distantly}, and further filtered by human experts. The links are constructed through open relation pattern mining \cite{li2018pattern} and manual selection. We compute word2vec \cite{mikolov2013distributed} on all PubMed papers and aggregate the word embeddings to get 200-dim features for all types of nodes. We further label a relatively small portion of diseases into eight categories. Each labeled disease has only one label.

The four datasets we prepare are from four different domains, which have been individually studied in some existing works on HNE. Among them, DBLP has been most commonly used, because all information about authors, papers, \etc. is public and there is no privacy issue, and the results are often more interpretable to researchers in computer science related domains. Other real-world networks like Yelp and IMDB have been commonly studied for recommender systems. These networks are naturally heterogeneous, including at least users and items (\eg, businesses and movies), as well as some additional item descriptors (\eg, categories of businesses and genres of movies). Freebase is one of the most popular open-source knowledge graph, which is relatively smaller but cleaner compared with the others (\eg, YAGO \cite{suchanek2007yago} and Wikidata \cite{vrandevcic2014wikidata}), where most entity and relation types are well defined. One major difference between conventional heterogeneous networks and knowledge graphs is the number of types of nodes and links. We further restrict the types of entities and relations inside Freebase, so as to get a heterogeneous network that is closer to a knowledge graph, while in the meantime does not have too many types of nodes and links. Therefore, most conventional HNE algorithms can be applied on this dataset and properly compared against the KB embedding ones. PubMed is a novel biomedical network we directly construct through text mining and manual processing on biomedical literature. This is the first time we make it available to the public, and we hope it to serve both the evaluation of HNE algorithms and novel downstream tasks in biomedical science such as biomedical information retrieval and disease evolution study.

We notice that several other heterogeneous network datasets such as OAG \cite{zhang2019oag, hu2020heterogeneous, dong2020HeterNRL} and IMDB \cite{wang2019heterogeneous, fu2020magnn, zhu2020hgcn} have been constructed recently in parallel to this work. Due to their similar nature and organization as some of our datasets (\eg, DBLP, Yelp), our pipeline can be easily adopted on these datasets, so we do not copy them here.

\subsection{Structure Analysis}
A summary of the statistics on the four datasets is provided in Table \ref{tab:stats} and Figure \ref{fig:nodetype}. As can be observed, the datasets have different sizes (numbers of nodes and links) and heterogeneity (numbers and distributions of node/link types). Moreover, due to the nature of the data sources, DBLP and PubMed networks are attributed, whereas Yelp and Freebase networks are abundantly labeled. A combination of these four datasets thus allows researchers to flexibly start by testing an HNE algorithm in the most appropriate settings, and eventually complete an all-around evaluation over all settings.

We also provide detailed analysis regarding several most widely concerned properties of heterogeneous networks, \ie, degree distribution (Figure \ref{fig:degree}), clustering coefficient (Figure \ref{fig:cc}), and number of frequent meta-paths (Figure \ref{fig:metapath}). 
In particular, degree distribution is known to significantly influence the performance of HNE algorithms due to the widely used node sampling process, whereas clustering coefficient impacts HNE algorithms that utilize latent community structures. Moreover, since many HNE algorithms rely on meta-paths, the skewer distribution of meta-paths can bias towards algorithms using fewer meta-paths. 

As we can see, the properties we concern are rather different across the four datasets we prepare.
For example, there are tighter links and more labels in Yelp, while there are more types of nodes and links in Freebase;  compared with nodes in Freebase and PubMed which clearly follow the long-tail degree distribution, certain types of nodes in DBLP and Yelp are always well connected (\eg, phrases in DBLP and businesses in Yelp), forming more \textit{star-shaped} subgraphs; the type-wise clustering coefficients and meta-path distributions are the most skewed in DBLP and most balanced in PubMed. The set of four datasets together provide a comprehensive benchmark towards the robustness and generalizability of various HNE algorithms (as we will also see in Section \ref{sec:exp}.2).

\subsection{Settings, Tasks, and Metrics}
We mainly compare all 13 algorithms under the setting of unsupervised unattributed HNE over all datasets, where the essential goal is to preserve different types of edges in the heterogeneous networks. Moreover, for message-passing algorithms that are particularly designed for attributed and semi-supervised HNE, we also conduct additional experiments for them in the corresponding settings. Particularly, due to the nature of the datasets, we evaluate attributed HNE on DBLP and PubMed datasets where node attributes are available, and semi-supervised HNE on Yelp and Freebase where node labels are abundant. We always test the computed network embeddings on the two standard network mining tasks of node classification and link prediction. Note that, while most existing studies on novel HNE algorithms have been focusing on these two standard tasks \cite{dong2017metapath2vec, fu2017hin2vec, tang2015pte, shi2018easing, wang2019heterogeneous, fu2020magnn, hu2020heterogeneous}, we notice that there are also various other tasks due to the wide usage of heterogeneous networks in real-world applications \cite{yang2018did, liu2018interactive, zhang2019your, fan2019metapath, niu2020dual, luo2020dynamic, hong2020heteta}. While the performance of HNE there can be rather task-dependant and data-dependant, to firstly simplifying the task into either a standard node classification or link prediction problem can often serve to provide more insights into the task and dataset, which can help the further development of novel HNE algorithms.

For the standard unattributed unsupervised HNE setting, we first randomly hide 20\% links and train all HNE algorithms with the remaining 80\% links. 
For node classification, we then train a separate linear Support Vector Machine (LinearSVC) \cite{fan2008liblinear} based on the learned embeddings on 80\% of the labeled nodes and predict on the remaining 20\%. We repeat the process for five times and compute the average scores regarding macro-F1 (across all labels) and micro-F1 (across all nodes). 
For link prediction, we use the Hadamard function to construct feature vectors for node pairs, train a two-class LinearSVC on the 80\% training links and evaluate towards the 20\% held out links. We also repeat the process for five times and compute the two metrics of AUC (area under the ROC curve) and MRR (mean reciprocal rank). AUC is a standard measure for classification, where we regard link prediction as a binary classification problem, and MRR is a standard measure for ranking, where we regard link prediction as a link retrieval problem. Since exhaustive computation over all node pairs is too heavy, we always use the two-hop neighbors as the candidates for all nodes.
For attributed HNE, node features are used during the training of HNE algorithms, whereas for semi-supervised HNE, certain amounts of node labels are used (80\% by default).

\section {Experimental Evaluations} 
\label{sec:exp}

\begin{table*}[h]
 \centering
 \begin{tabular}{lcccccccc}
 \hline
\multirow{2}{*}{Model}&\multicolumn{4}{c}{Node classification (Macro-F1/Micro-F1)}&\multicolumn{4}{c}{Link prediction (AUC/MRR)}\\
\cline{2-9}
&DBLP&Yelp&Freebase&PubMed&DBLP&Yelp&Freebase&PubMed\\
\hline
metapath2vec & 43.85/55.07 & 5.16/23.32 & 20.55/46.43 & 12.90/15.51 & 65.26/90.68 & 80.52/99.72 & 56.14/78.24 & 69.38/84.79 \\
PTE & 43.34/54.53 & 5.10/23.24 & 10.25/39.87 & 09.74/12.27 & 57.72/77.51 & 50.32/68.84 & 57.89/78.23 & 70.36/89.54 \\
HIN2Vec & 12.17/25.88 & 5.12/23.25 &17.40/41.92  & 10.93/15.31 & 53.29/75.47 & 51.64/66.71 & 58.11/81.65 & 69.68/84.48 \\
AspEm & 33.07/43.85 & 5.40/23.82 &23.26/45.42 &11.19/14.44 & 67.20/91.46 & 76.10/95.18 & 55.80/77.70 & 68.31/87.43  \\
HEER & 09.72/27.72 & 5.03/22.92 & 12.96/37.51 & 11.73/15.29 & 53.00/72.76 & 73.72/95.92 & 55.78/78.31 & 69.06/88.42 \\
\hline
R-GCN & 09.38/13.39 & 5.10/23.24 & 06.89/38.02 & 10.75/12.73 & 50.50/73.35 & 72.17/97.46 & 50.18/74.01 & 63.33/81.19 \\
HAN & 07.91/16.98 & 5.10/23.24 & 06.90/38.01 & 09.54/12.18 & 50.24/73.10 & N/A & 51.50/74.13 & 65.85/85.33 \\
MAGNN & 06.74/10.35 & 5.10/23.24 & 06.89/38.02 & 10.30/12.60 & 50.10/73.26 & 50.03/69.81 & 50.12/74.18 & 61.11/90.01 \\
HGT & 15.17/32.05 & 5.07/23.12 & 23.06/46.51 & 11.24/18.72 & 59.98/83.13 & 79.00/99.66 & 55.68/79.46 & 73.00/88.05 \\
\hline
TransE & 22.76/37.18 & 5.05/23.03 & 31.83/52.04 & 11.40/15.16 & 63.53/86.29 & 69.13/83.66 & 52.84/75.80 & 67.95/84.69 \\
DistMult & 11.42/25.07 & 5.04/23.00 & 23.82/45.50 & 11.27/15.79 & 52.87/74.84 & 80.28/99.73 & 54.91/78.04 & 70.61/90.64 \\
ComplEx & 20.48/37.34 & 5.05/23.03 & 35.26/52.03 & 09.84/18.51 & 65.92/90.01 & 80.11/99.73 & 60.43/84.22 & 75.96/92.47 \\
ConvE & 12.42/26.42 & 5.09/23.02 & 24.57/47.61 & 13.00/14.49 & 54.03/75.31 & 78.55/99.70 & 54.29/76.11 & 71.81/89.82 \\
\hline
\end{tabular}
 \caption{\label{tab:perform}\textbf{Performance comparison (\%) under the standard setting of unattributed unsupervised HNE. }}
\end{table*}

\begin{figure*}[t!]
\centering
\hspace{-5pt}
\subfigure{
\includegraphics[width=0.24\textwidth]{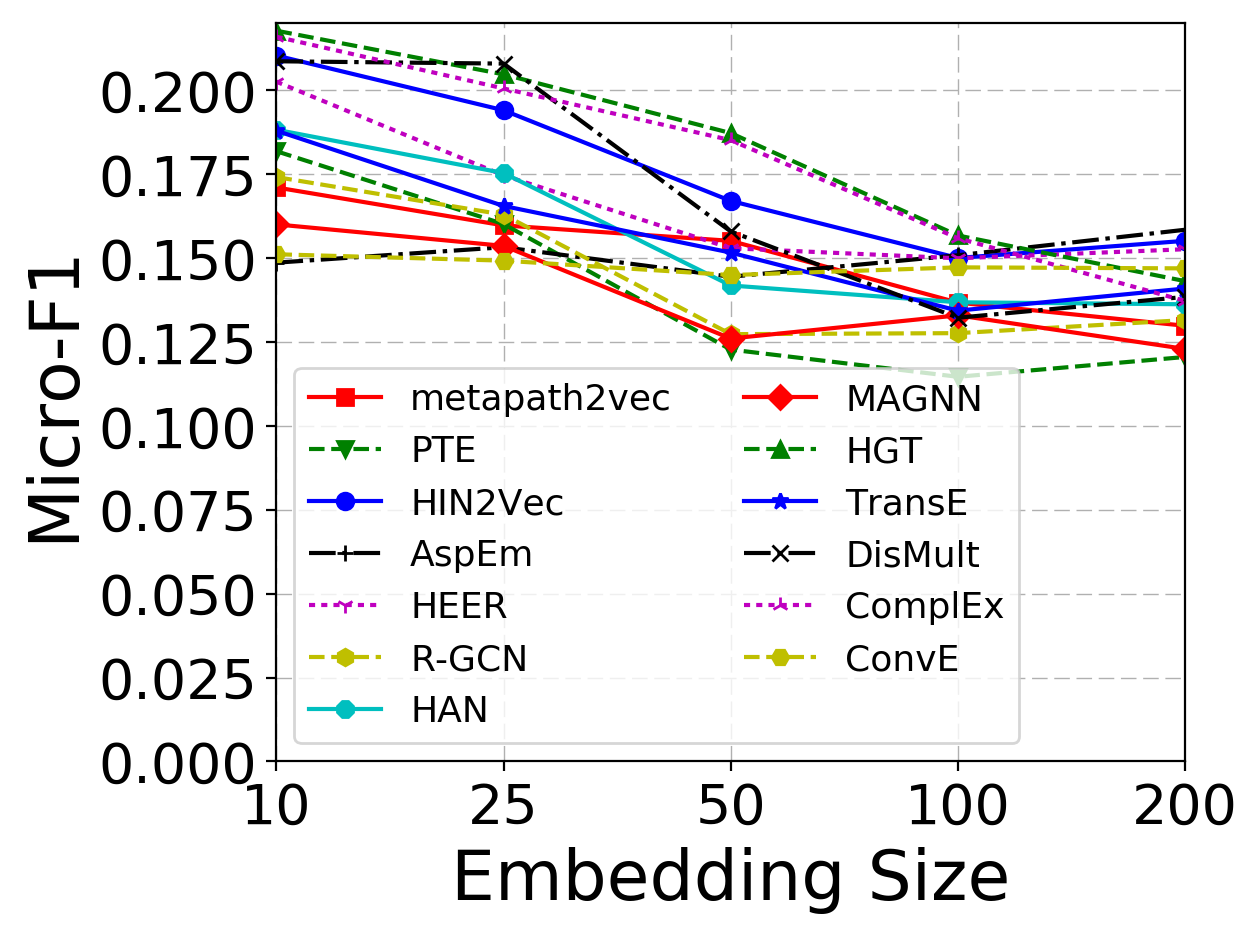}}
\hspace{-5pt}
\subfigure{
\includegraphics[width=0.24\textwidth]{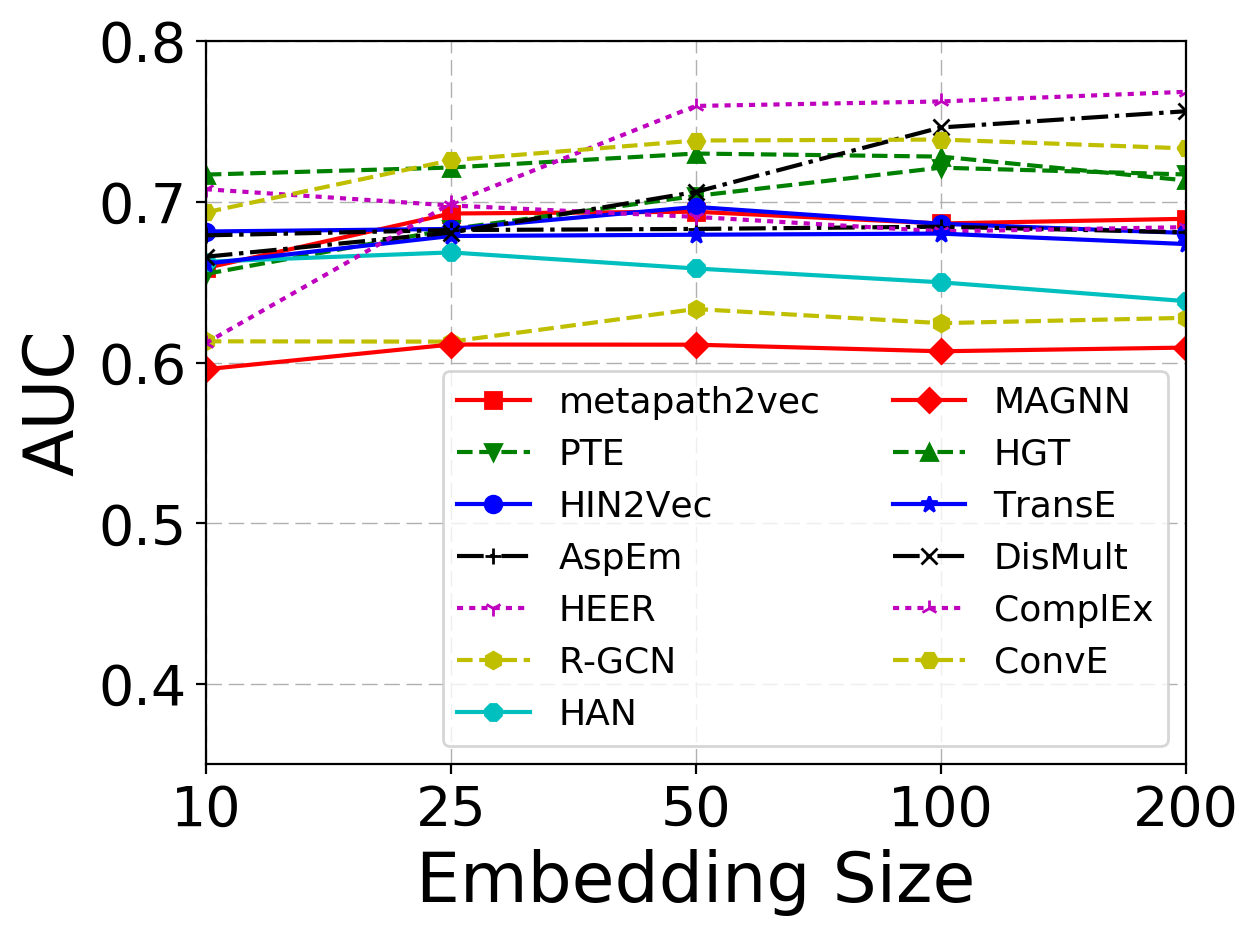}}
\hspace{-5pt}
\subfigure{
\includegraphics[width=0.24\textwidth]{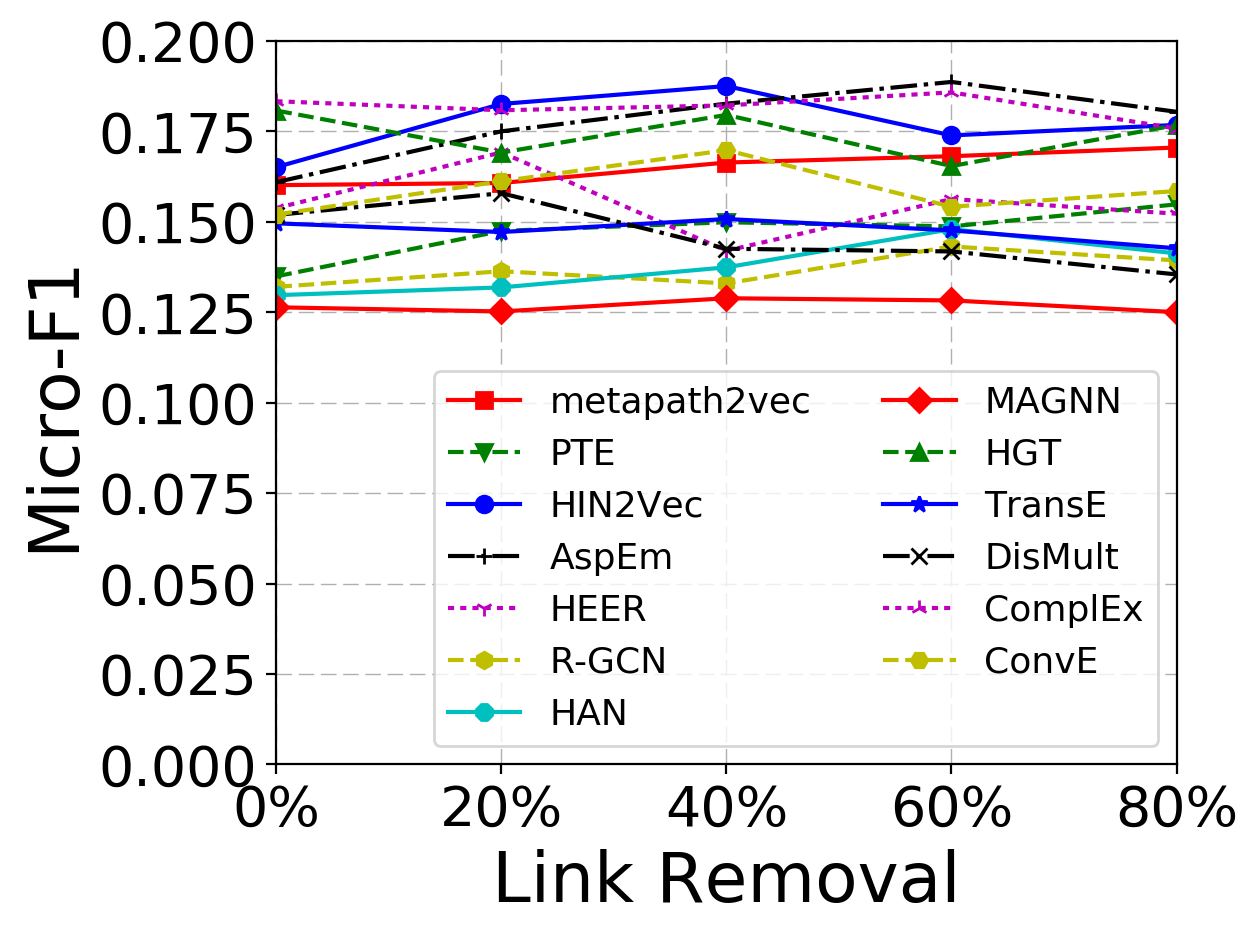}}
\hspace{-5pt}
\subfigure{
\includegraphics[width=0.24\textwidth]{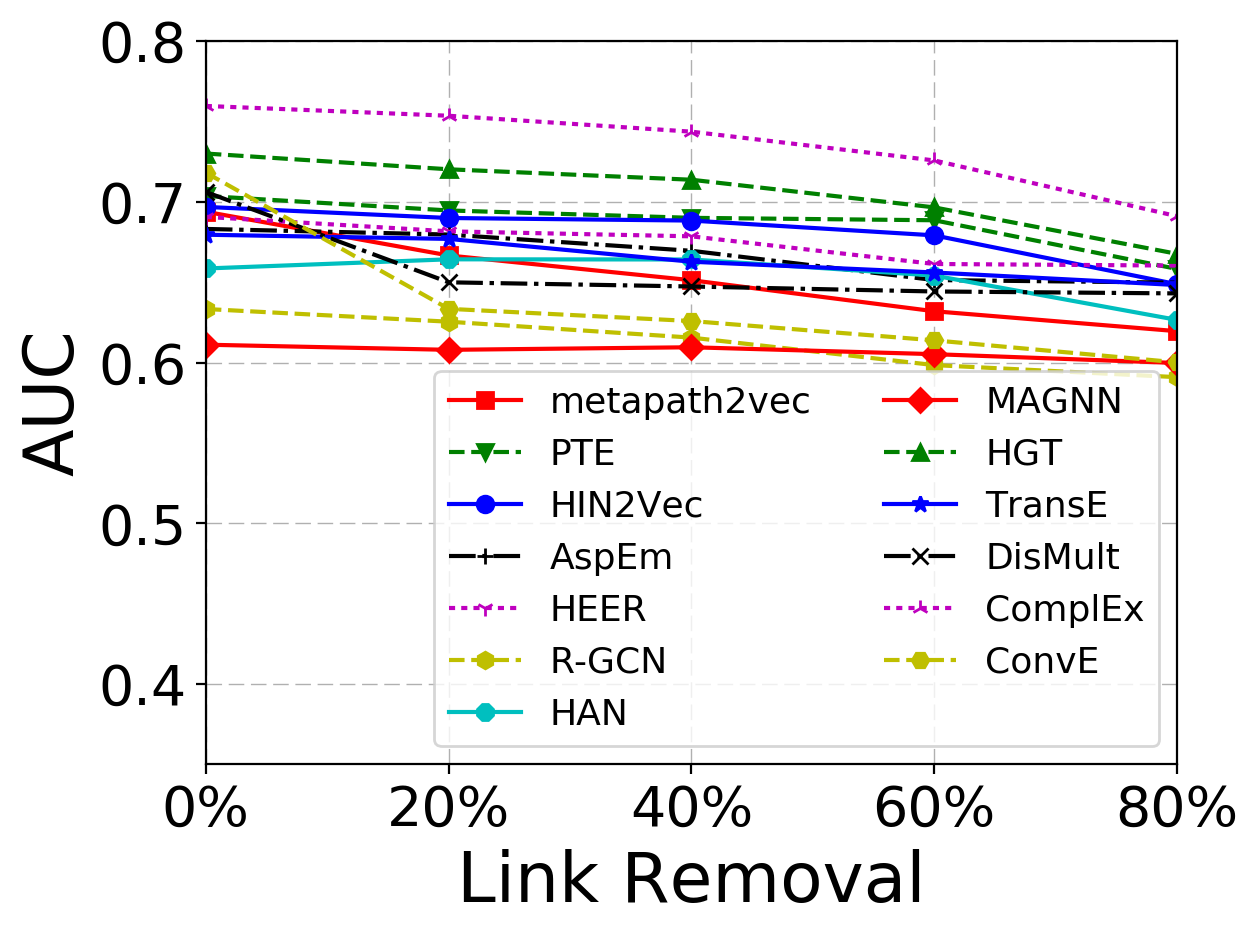}}
\hspace{-5pt}
\caption{\textbf{Performance comparison under controlled experiments with varying emb.~sizes and link removals (PubMed).}}
\label{fig:control}
\end{figure*}

\begin{figure*}[t!]
\centering
\hspace{-5pt}
\subfigure{
\includegraphics[width=0.24\textwidth]{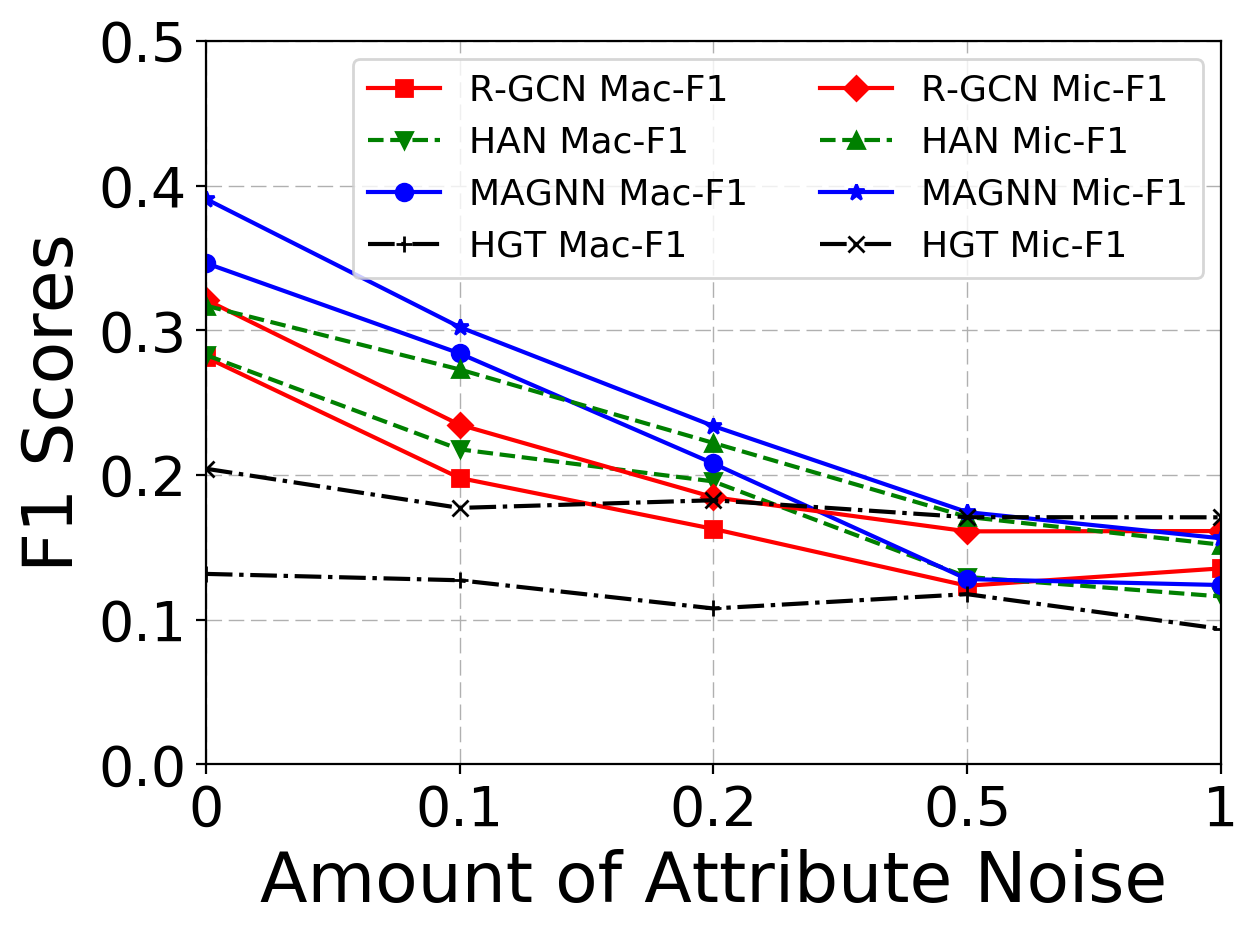}}
\hspace{-5pt}
\subfigure{
\includegraphics[width=0.24\textwidth]{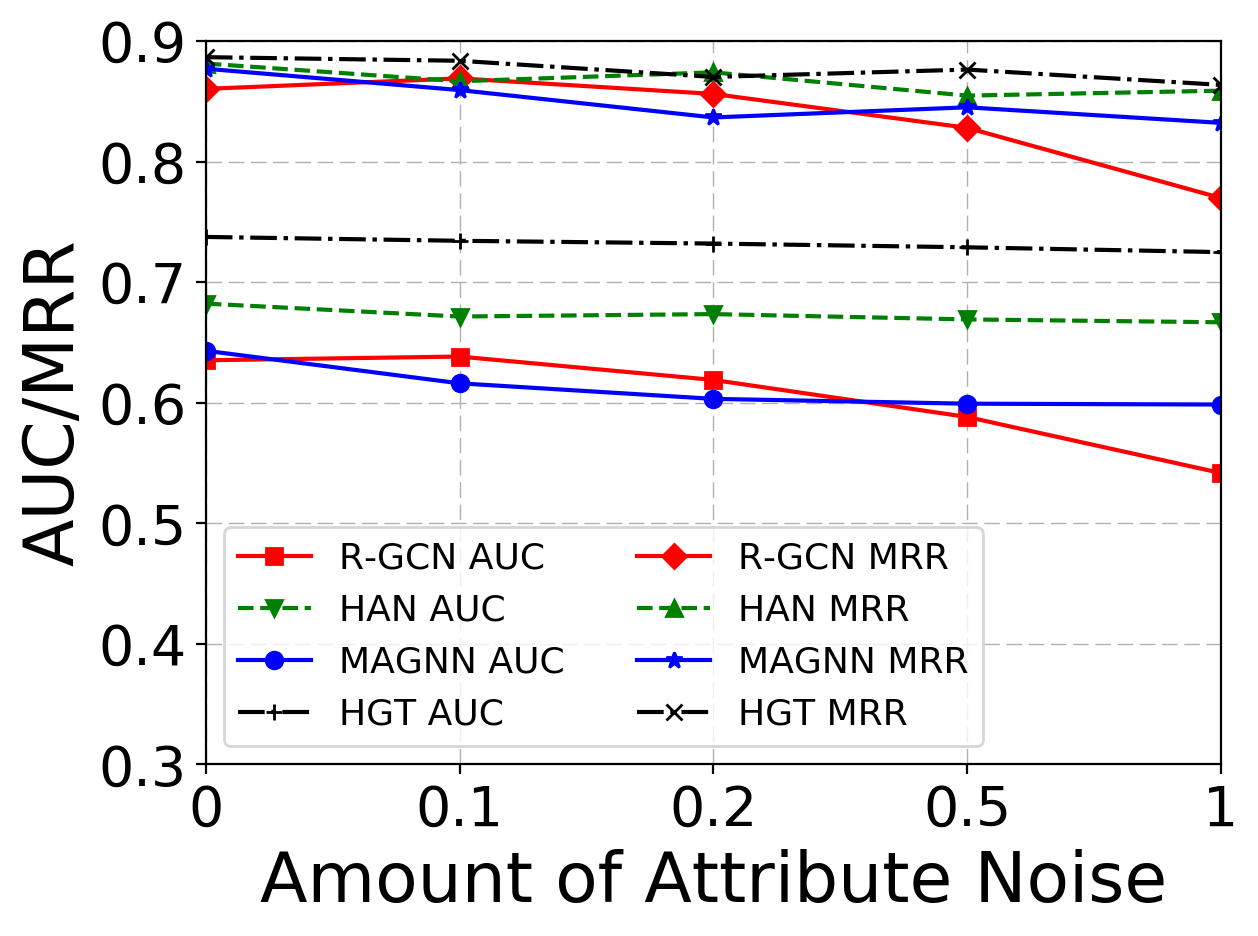}}
\hspace{-5pt}
\subfigure{
\includegraphics[width=0.24\textwidth]{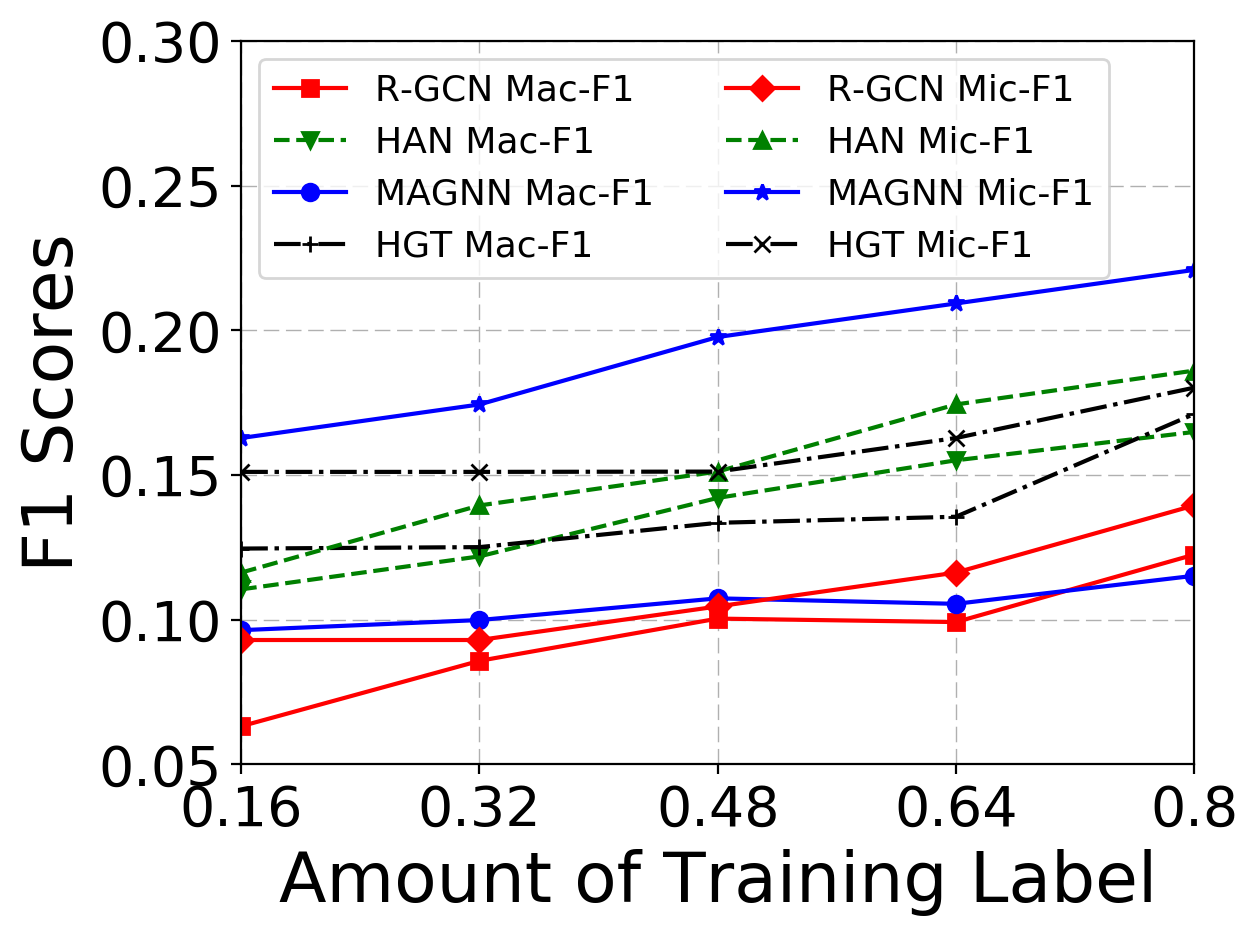}}
\hspace{-5pt}
\subfigure{
\includegraphics[width=0.24\textwidth]{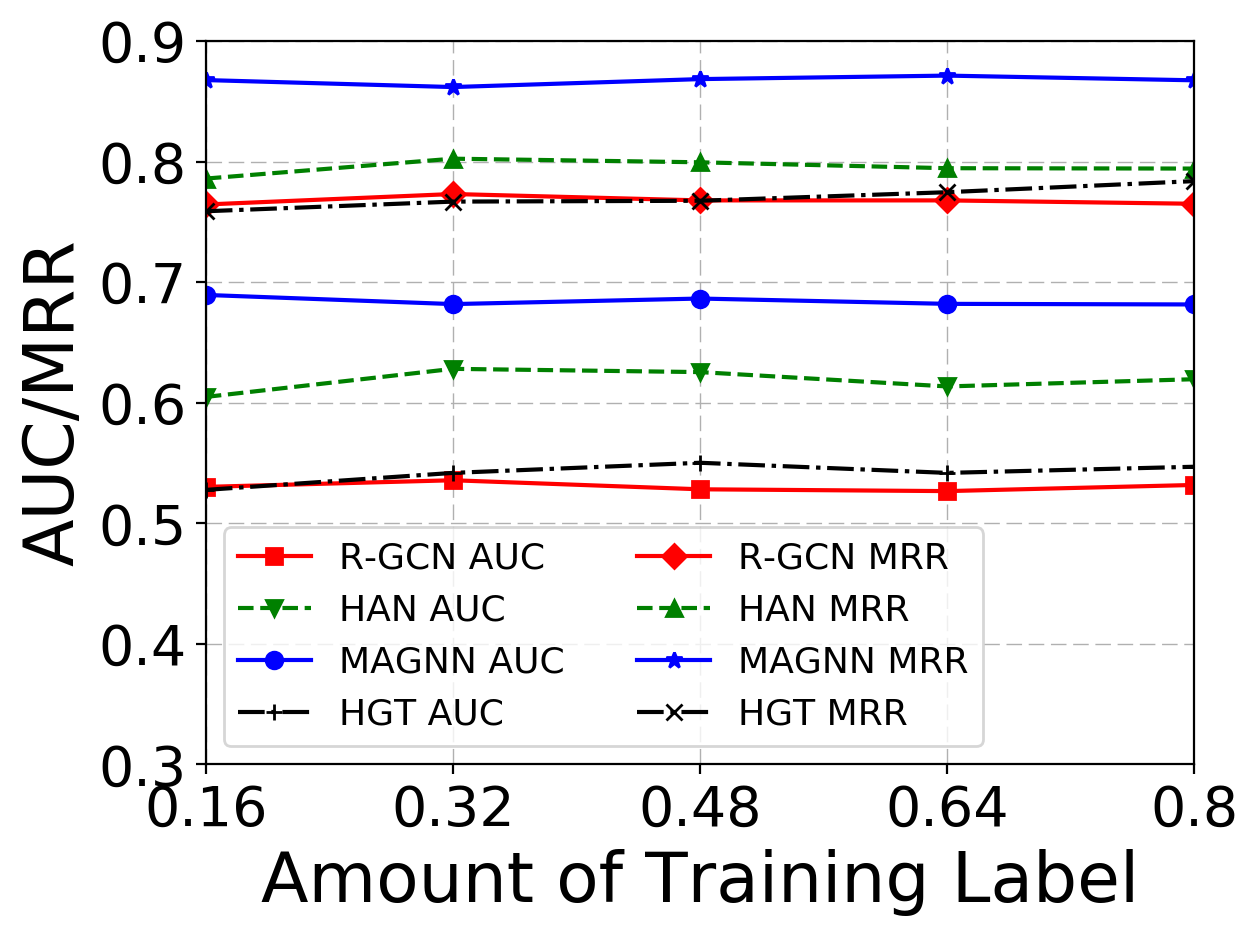}}
\hspace{-5pt}
\caption{\textbf{Performance comparison under controlled experiments with varying label amount and attr.~noise (PubMed).}}
\label{fig:label}
\end{figure*}

\subsection{Algorithms and Modifications}
We amend the implementations of 13 popular HNE algorithms for seamless and efficient experimental evaluations on our prepared datasets. The algorithms we choose and the modifications we make are as follows.
\begin{itemize}
\item \textbf{metapath2vec} \cite{dong2017metapath2vec}: Since the original implementation contains a large amount of hard-coded data-specific settings such as node types and meta-paths, and the optimization is unstable and limited as it only examines one type of meta-path based context, we completely reimplement the algorithm. In particular, we first run random walks to learn the weights of different meta-paths based on the number of sampled instances, and then train the model using the unified loss function, which is a weighted sum over the loss functions of individual meta-paths. Both the random walk and meta-path-based embedding optimization are implemented with multi-threads in parallel.
\item \textbf{PTE} \cite{tang2015pte}: Instead of accepting labeled texts as input and working on text networks with the specific three types of nodes (word, document, and label) and three types of links (word-word, document-word, and label-word), we revise the original implementation and allow the model to consume heterogeneous networks directly with arbitrary types of nodes and links by adding more type-specific objective functions.
\item \textbf{HIN2Vec} \cite{fu2017hin2vec}: We remove unnecessary data preprocessing codes and modify the original implementation so that the program first generates random walks, then trains the model, and finally outputs node embeddings only.
\item \textbf{AspEm} \cite{shi2018aspem}: We clean up the hard-coded data-specific settings in the original implementation and write a script to connect the different components of automatically selecting the aspects with the least incompatibilities, as well as learning, matching, and concatenating the embeddings based on different aspects.
\item \textbf{HEER} \cite{shi2018easing}: We remove the hard-coded data-specific settings and largely simplify the data preprocessing step in the original implementation by skipping the knockout step and disentangling the graph building step. 
\item \textbf{R-GCN} \cite{schlichtkrull2018modeling}: The existing implementation from DGL \cite{wang2019dgl} is only scalable to heterogeneous networks with thousands of nodes, due to the requirement of putting the whole graphs into memory during graph convolutions. To scale up R-GCN, we perform fixed-sized node and link sampling for batch-wise training following the framework of GraphSAGE \cite{hamilton2017inductive}. 
\item \textbf{HAN} \cite{wang2019heterogeneous}: Since the original implementation of HAN contains a large amount of hard-coded data-specific settings such as node types and meta-paths, and is unfeasible for large-scale datasets due to the same reason as R-GCN, we completely reimplement the HAN algorithm based on our implementation of R-GCN. In particular, we first automatically construct meta-path based adjacency lists for the chosen node type, and then sample the neighborhood for the seed nodes during batch-wise training. 
\item \textbf{MAGNN} \cite{fu2020magnn}: The original DGL-based \cite{wang2019dgl} unsupervised implementation of MAGNN solely looks at predicting the links between users and artists in a music website dataset with the use of a single neural layer. Hence, we remove all the hard-coded data-specific settings and refactor the entire pipeline so that the model can support multi-layer mini-batch training over arbitrary link types.
\item \textbf{HGT} \cite{hu2020heterogeneous}: The original PyG-based \cite{Fey/Lenssen/2019} implementation of HGT targets on the specific task of author disambiguation among the associated papers in a dynamic academic graph \cite{zhang2019oag}. Therefore, we refactor it by removing hard-coded data-specific settings, assigning the same timestamp to all the nodes, and conducting training over all types of links. 
\item \textbf{TransE} \cite{bordes2013translating}: We modify the OpenKE \cite{han2018openke} implementation so that the model outputs node embeddings only.
\item \textbf{DistMult} \cite{yang2014embedding}: We remove the hard-coded data-specific settings and largely simplify the data preprocessing step in the original implementation.
\item \textbf{ComplEx} \cite{trouillon2016complex}: Same as for TransE.
\item \textbf{ConvE} \cite{dettmers2018convolutional}: Same as for DistMult.
\end{itemize}

We set the embedding size of all algorithms to 50 by default, and tune other hyperparameters following the original papers through standard five-fold cross validation on all datasets.
We have put the implementation of all compared algorithms in a python package and released them together with the datasets to constitute an open-source ready-to-use HNE benchmark.\textsuperscript{\ref{repo}}

\subsection{Performance Benchmarks}
We provide systematic experimental comparisons of the 13 popular state-of-the-art HNE algorithms across our four datasets, on the scenarios of unsupervised unattributed HNE, attributed HNE, and semi-supervised HNE.

Table \ref{tab:perform} shows the performance of compared algorithms on unsupervised unattributed HNE, evaluated towards node classification and link prediction. We have the following observations. 

From the perspective of compared algorithms: 

(1) Proximity-preserving algorithms often perform well on both tasks under the unsupervised unattributed HNE setting, which explains why proximity-preserving is the most widely used HNE or even general network embedding framework when node attributes and labels are unavailable. Among the proximity-preserving methods, HIN2Vec and HEER show reasonable results on link prediction but perform not so well on node classification (especially on DBLP and Freebase). In fact, these two methods focus on modeling link representations in their objectives ($\bmA_{\mathcal{M}}$ in HIN2Vec and $\bmA_l$ in HEER), thus are more suitable for link prediction. 

(2) Under the unsupervised unattributed HNE setting, message-passing methods perform poorly except for HGT, especially on node classification. As we discuss before, message-passing methods are known to excel due to their integration of node attributes, link structures, and training labels. When neither of node attributes and labels are available, we use random vectors as node features and adopt a link prediction loss, which largely limits the performance of R-GCN, HAN, and MAGNN. We will focus our evaluation on the message-passing algorithms in the attributed and semi-supervised HNE settings later. 
On the contrary, HGT exhibits competitive results on both node classification and link prediction. This is attributed to the usage of node-type and link-type dependent parameters which maintains dedicated representations for different types of nodes and links. In addition, the heterogeneous mini-batch graph sampling algorithm designed by HGT further reduces the loss of structural information due to sampling and boosts the performance to a greater extent.
Finally, the link prediction result of HAN on the Yelp dataset is not available. This is because HAN can only embed one type of nodes at a time (we embed Business in Yelp) and thus predict the link between two nodes with the same type (\ie, Business-Business). However, all links in Yelp connect distinct types of nodes (\eg, Business-Location, Business-User), and HAN cannot predict such links (thus marked as N/A).

(3) Relation-learning methods such as TransE and ComplEx perform better on Freebase and PubMed on both tasks, especially on link prediction. In fact, in Table \ref{tab:stats} and Figure \ref{fig:nodetype} we can observe that both datasets (especially Freebase) have more link types. Relation-learning approaches, which are mainly designed to embed knowledge graphs (\eg, Freebase), can better capture the semantics of numerous types of direct links. 

From the perspective of datasets:

(1) All approaches have relatively low F1 scores on Yelp and PubMed (especially Yelp) on node classification. This is because both datasets have larger numbers of classes (\ie, 16 in Yelp and 8 in PubMed) as shown in Table \ref{tab:stats}. Moreover, unlike the cases of the other datasets, a node in Yelp can have multiple labels, which makes the classification task more challenging. 

(2) In Figure \ref{fig:degree}, we can observe that the degree distribution of Freebase is more skewed. Therefore, when we conduct link sampling or random walks on Freebase during representation learning, nodes with lower degrees will be sampled less frequently and their representations may not be learned accurately. This observation may explain why the link prediction metrics on Freebase are in general lower than those on Yelp and PubMed.

(3) As we can see in Figure \ref{fig:nodetype}-\ref{fig:metapath}, most studied network properties are more balanced on Freebase and PubMed (especially PubMed) across different types of nodes and links. This in general makes both the node classification and link prediction tasks harder for all algorithms, and also makes the gaps among different algorithms smaller.

\subsection{Ablation Studies}
To provide an in-depth performance comparison among various HNE algorithms, we further conduct controlled experiments by varying the embedding sizes and randomly removing links from the training set. 

In Figure \ref{fig:control}, we show the micro-F1 scores for node classification and AUC scores for link prediction computed on the PubMed dataset. We omit the other results here, which can be easily computed in our provided benchmark package.
As we can observe, some algorithms are more robust to varying settings while some others are more sensitive. In general, varying embedding size and link removal can significantly impact the performance of most algorithms on both tasks, and sometimes can even lead to different ordering of certain algorithms. This again emphasizes the importance of setting up standard benchmark including datasets and evaluation protocols for systematic HNE algorithm evaluation.
In particular, on PubMed, larger embedding sizes like over 50 can harm the performance of most algorithms especially on node classification, probably due to overfitting with the limited labeled data. 
Interestingly, the random removal of links does have a negative impact on link prediction, but it does not necessarily harm node classification. This means that node classes and link structures may not always be tightly correlated, and even parts of the links already provide the necessary information useful enough for node classification.

Towards the evaluation of the message-passing HNE algorithms designed to integrate node attributes and labels into representation learning like R-GCN, HAN, MAGNN, and HGT, we also conduct controlled experiments by adding random Gaussian noises to the node attributes and masking different amounts of training labels.

In Figure \ref{fig:label}, we show the results on the PubMed dataset. As we can observe, the scores in most subfigures are significantly higher than the scores in Table \ref{tab:perform}, indicating the effectiveness of R-GCN, HAN, MAGNN, and HGT in integrating node attributes and labels for HNE. 
In particular, the incorporation of node attributes boosts the node classification results of R-GCN, HAN, and MAGNN significantly (almost tripling the F1 scores and significantly higher than all algorithms that do not use node attributes), but it offers very little help to HGT. This suggests that R-GCN, HAN, and MAGNN can effectively leverage the semantic information associated with attributes, whereas HGT relies more on the network structures and type information of nodes and links. Moreover, MAGNN achieves the highest node classification results with large margins as it successfully incorporates the attributes of intermediate nodes along the meta-paths (which are ignored by HAN).
In addition, when random noises with larger variances are added to node attributes, the performance of node classification significantly drops, while the performance of link prediction is less affected. 
As more training labels become available, without a surprise, the node classification results of all four algorithms increase, but surprisingly the link prediction results are almost not affected. 
These observations again reveal the different natures of the two tasks, where node classes are more related to node contents, whereas links should be typically inferred from structural information.

\section{Future}
\label{sec:future}
In this work, we present a comprehensive survey on various existing HNE algorithms, and provide benchmark datasets and baseline implementations to ease future research in this direction. While HNE has already demonstrated strong performance across a variety of downstream tasks, it is still in its infancy with many open challenges. To conclude this work and inspire future research, we now briefly discuss the limitation of current HNE and several specific directions potentially worth pursuing. 

\vspace{-3pt}
\header{Beyond homophily.}
As we formulate in Eq.~(\ref{eq:general}), current HNE algorithms focus on the leverage of network homophily. Due to recent research on homogeneous networks that study the combination of \textit{positional} and \textit{structural} embedding \cite{you2019position, ribeiro2017struc2vec}, it would be interesting to explore how to generalize such design principles and paradigms to HNE. Particularly, in heterogeneous networks, relative positions and structural roles of nodes can both be measured under different meta-paths or meta-graphs, which are naturally more informative and diverse. However, such considerations also introduce harder computational challenges.

\vspace{-3pt}
\header{Beyond accuracy.}
Most, if not all, existing research on HNE has primarily focused on the accuracy towards different downstream tasks. It would be interesting to further study the \textit{scalability and efficiency} (for large-scale networks) \cite{hamilton2017inductive, hu2019strategies}, \textit{temporal adaptability} (for dynamic evolving networks) \cite{zhou2018dynamic, du2018dynamic}, \textit{robustness} (towards adversarial attacks) \cite{wang2018graphgan, zugner2018adversarial}, \textit{interpretability} \cite{liu2018interpretation}, \textit{uncertainty} \cite{chen2019embedding}, \textit{fairness} \cite{rahman2019fairwalk, bose2019compositional} of HNE, and so on.
 
\vspace{-3pt}
\header{Beyond node embedding.}
Graph- and subgraph-level embeddings have been intensively studied on homogeneous networks to enable graph-level classification \cite{xu2018powerful, ying2018hierarchical} and unsupervised message-passing model training \cite{velivckovic2018deep, sun2019infograph}, but they are hardly studied on heterogeneous networks \cite{park2020deep}. Although existing works like HIN2Vec \cite{fu2017hin2vec} study the embedding of meta-paths to improve the embedding of nodes, direct applications of graph- and subgraph-level embeddings in the context of heterogeneous networks largely remain nascent. 
 
\vspace{-3pt}
\header{Revisiting KB embedding.}
The difference between KB embedding and other types of HNE is mainly due to the numbers of node and link types. Direct application of KB embedding to heterogeneous networks fails to consider meta-paths with rich semantics, whereas directly applying HNE to KB is unrealistic due to the exponential number of meta-paths. However, it would still be interesting to study the intersection between these two groups of methods (as well as two types of data). For example, how can we combine the ideas of meta-paths on heterogeneous networks and embedding transformation on KB for HNE with more semantic-aware transformations? How can we devise truncated random walk based methods for KB embedding to include higher-order relations?
 
 \vspace{-3pt}
\header{Modeling heterogeneous contexts.}
Heterogeneous networks mainly model different types of nodes and links. However, networks nowadays are often associated with rich contents, which provide contexts of the nodes, links, and subnetworks \cite{wang2019mcne, yang2020multisage, yang2019cubenet}. There have been studies exploiting text-rich heterogeneous information networks for taxonomy construction \cite{shi2019discovering,shang2020nettaxo} and text classification \cite{zhang2019higitclass,zhang2020minimally}. Meanwhile, how to model heterogeneous interactions under multi-faceted contexts through the integration of multi-modal content and structure could be a challenging but rewarding research area.

\vspace{-5pt}
\header{Understanding the limitation.}
While HNE (as well as many neural representation learning models) has demonstrated strong performance in various domains, it is worthwhile to understand its potential limits. For example, when do modern HNE algorithms work better compared with traditional network mining approaches (\eg, path counting, subgraph matching, non-neural or linear propagation)? How can we join the advantages of both worlds?
Moreover, while there has been intensive research on the mathematical mechanisms behind neural networks for homogeneous network data (\eg, smoothing, low-pass filtering, invariant and equivariant transformations), by unifying existing models on HNE, this work also aims to stimulate further theoretical studies on the power and limitation of HNE.

\section*{Acknowledgement}
Special thanks to Dr. Hanghang Tong for his generous help in the polish of this work. Research was sponsored in part by the U.S.~ Army Research Laboratory under Cooperative Agreement No.~W911NF-09-2-0053 and W911NF-13-1-0193, DARPA No.~W911NF-17-C-0099 and FA8750-19-2-1004, National Science Foundation IIS 16-18481, IIS 17-04532, and IIS-17-41317, DTRA HDTRA11810026, and grant 1U54GM114838 awarded by NIGMS through funds provided by the trans-NIH Big Data to Knowledge (BD2K) initiative.
\bibliographystyle{IEEEtran}
\bibliography{carlyang} 
\newpage
\begin{IEEEbiography}[{\includegraphics[width=1in,
height=1.25in,clip,keepaspectratio]{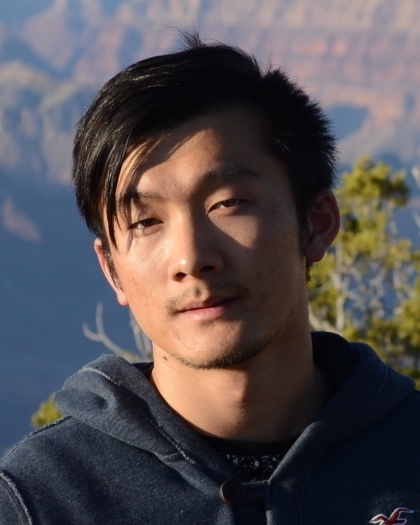}}]{Carl Yang} received his Ph.D.~under Jiawei Han in Computer Science at University of Illinois, Urbana-Champaign. He received his B.Eng.~in Computer Science and Engineering at Zhejiang University in 2014. In his research, he develops data-driven techniques and neural architectures for learning with context-rich network data. Carl's research results have been published in top conferences like KDD, NeurIPS, WWW, ICDE, SIGIR, ICML. Carl also received the Dissertation Completion Fellowship of UIUC in 2020, Best Paper Award of ICDM 2020, and has started as an Assistant Professor in Emory University after Ph.D.~graduation.
\end{IEEEbiography}
\vspace{-40pt}
\begin{IEEEbiography}[{\includegraphics[width=1in,height=1.25in,clip,keepaspectratio]{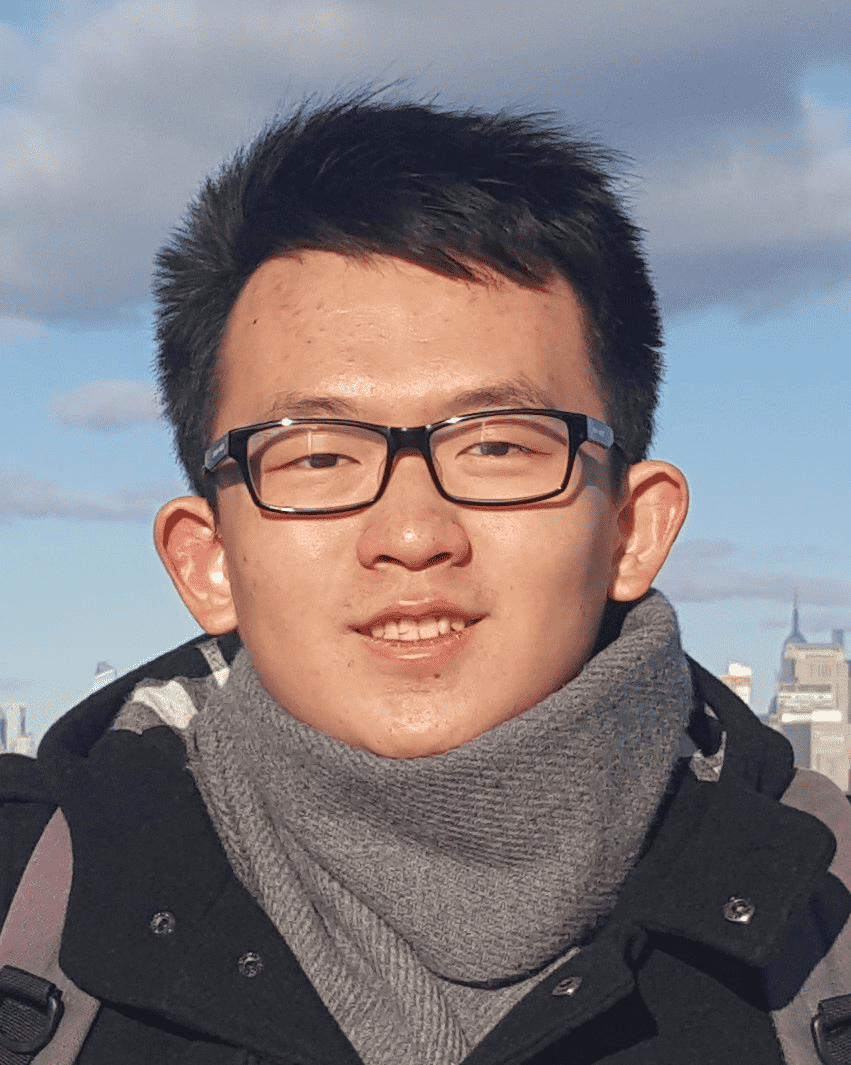}}]{Yuxin Xiao} is a master student at Carnegie Mellon University. He received his B.S.~in Computer Science and B.S.~in Statistics and Mathematics at the University of Illinois at Urbana-Champaign in 2020. His research focuses on machine learning on graph data and he has published first-authored papers in WWW and IEEE BigData. Yuxin also received the CRA Outstanding Undergraduate Researcher Award and C.W.~Gear Outstanding Undergraduate Award at UIUC in 2020.
\end{IEEEbiography}
\vspace{-40pt}
\begin{IEEEbiography}[{\includegraphics[width=1in,height=1.25in,clip,keepaspectratio]{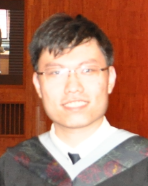}}]{Yu Zhang} is a Ph.D.~candidate in computer science at the University of Illinois at Urbana-Champaign, advised by Prof. Jiawei Han. He received his M.S. degree at UIUC in 2019 and his B.S. degree at Peking University in 2017. His research interests include text-rich network mining, text classification, and their applications to bioinformatics. He has published first-authored papers in SIGIR, WSDM, ICDM, and CIKM. Yu also received WWW Best Poster Award Honorable Mention in 2018.
\end{IEEEbiography}
\vspace{-40pt}
\begin{IEEEbiography}[{\includegraphics[width=1in,height=1.25in,clip,keepaspectratio]{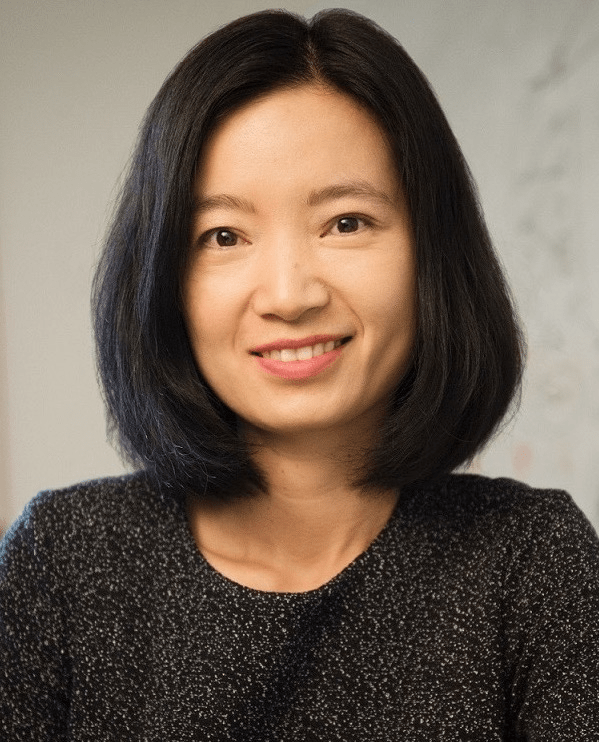}}]{Yizhou Sun} is an associate professor at department of computer science of UCLA. She received her Ph.D.~in Computer Science from the University of Illinois at Urbana-Champaign in 2012. Her principal research interest is on mining graphs/networks, and more generally in data mining, machine learning, and network science, with a focus on modeling novel problems and proposing scalable algorithms for large-scale, real-world applications. She is a pioneer researcher in mining heterogeneous information network, with a recent focus on deep learning on graphs/networks. Yizhou has over 100 publications in books, journals, and major conferences. Tutorials of her research have been given in many premier conferences. She received 2012 ACM SIGKDD Best Student Paper Award, 2013 ACM SIGKDD Doctoral Dissertation Award, 2013 Yahoo ACE (Academic Career Enhancement) Award, 2015 NSF CAREER Award, 2016 CS@ILLINOIS Distinguished Educator Award, 2018 Amazon Research Award, and 2019 Okawa Foundation Research Grant.
\end{IEEEbiography}
\vspace{-40pt}
\begin{IEEEbiography}[{\includegraphics[width=1in,height=1.25in,clip,keepaspectratio]{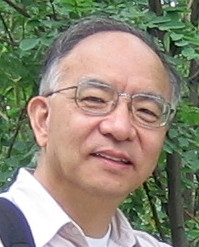}}]{Jiawei Han} is Michael Aiken Chair Professor in the Department of Computer Science, University
of Illinois at Urbana-Champaign. He has been researching into data mining, information network analysis, and database systems, and their various applications, with over 600 publications. He served as the founding Editor-in-Chief of ACM Transactions on Knowledge Discovery from Data (TKDD) (2007-2012). Jiawei has received ACM SIGKDD Innovation Award (2004), IEEE 
Computer Society Technical Achievement Award (2005), IEEE Computer Society W.~Wallace McDowell 
Award (2009), Daniel C.~Drucker Eminent Faculty Award at UIUC (2011), and Japan's Funai Achievement Award (2018). He is Fellow of ACM and Fellow of IEEE and served as co-Director of KnowEnG, a Center of Excellence in Big Data Computing (2014-2019), 
funded by NIH Big Data to Knowledge (BD2K) Initiative and as the Director of Information Network 
Academic Research Center (INARC) (2009-2016) supported by the Network Science-Collaborative 
Technology Alliance (NS-CTA) program of U.S.~Army Research Lab. His co-authored textbook ``Data Mining: Concepts and Techniques'' (Morgan Kaufmann) has been adopted popularly as a textbook worldwide.
\end{IEEEbiography}
\end{document}